%% file: transactions_on_computers_rePgbr.tex
\documentclass[10pt,journal,cspaper,compsoc]{IEEEtran}
\usepackage{ifpdf}

\ifCLASSOPTIONcompsoc
  \usepackage[nocompress]{cite}
\else
  \usepackage{cite}
\fi
\ifCLASSINFOpdf
  \usepackage[pdftex]{graphicx}
  \DeclareGraphicsExtensions{.pdf,.jpeg,.png}
\else
  \usepackage[dvips]{graphicx}
  \graphicspath{{./picture/eps/}}
  \DeclareGraphicsExtensions{.eps}
\fi
\usepackage[cmex10]{amsmath}
\newtheorem{mydef}{Calculation}
\usepackage{multirow}
\usepackage{algorithm}
\usepackage{algorithmic}
\ifCLASSOPTIONcompsoc
 \usepackage[caption=false,font=normalsize,labelfont=sf,textfont=sf]{subfig}
\else
 \usepackage[caption=false,font=footnotesize]{subfig}
\fi
\usepackage{fixltx2e}

\usepackage{tikz}

\newcommand{\kW}[1]{$#1~kW$}
\newcommand{\rePGBR}{$re$PGBR}
\newlength{\figurewidth}
\begin{document}

\title{Technical Report -- Renewable Energy-Aware Information-Centric Networking}

\author{Julien~Mineraud,
	Liang~Wang,
        Sasitharan~Balasubramaniam,~\IEEEmembership{Senior~Member,~IEEE}
        and~Jussi~Kangasharju,~\IEEEmembership{Member,~IEEE}
\IEEEcompsocitemizethanks{\IEEEcompsocthanksitem J. Mineraud, L. Wang and J. Kangasharju are with the Department
of Computer Science, University of Helsinki, Finland.\protect\\
E-mail: firstname.lastname@cs.helsinki.fi.
\IEEEcompsocthanksitem S. Balasubramaniam is with the Department of Electronic and Communication Engineering, Tampere University of Technology, Tampere, Finland.\protect\\
E-mail: sasi.bala@tut.fi}
\thanks{}}

%
%

\markboth{Technical Report in University of Helsinki, Finland, 2014-01-01}%
{Mineraud \MakeLowercase{\textit{et al.}}: Reducing Data Center Energy Use Via Renewable Energy Aware Routing and In-Network Caching}

\input{abstract}

\maketitle

\IEEEdisplaynotcompsoctitleabstractindextext

%
\IEEEpeerreviewmaketitle

\ifCLASSOPTIONcompsoc
 \noindent\raisebox{2\baselineskip}[0pt][0pt]%
 {\parbox{\columnwidth}{\section{Introduction}\label{sec:introduction}%
 \global\everypar=\everypar}}%
 \vspace{-1\baselineskip}\vspace{-\parskip}\par
\else
 \section{Introduction}\label{sec:introduction}\par
\fi
\input{introduction}

\section{Related work}\label{sec:relatedWork}
In this section, we will present related work in green Internet and information-centric caching, as these are most closely related to our proposed solution.

\subsection{Greening the Internet}
\input{relatedWork_routing}

\subsection{Content Caching Strategies}
\input{relatedWork_caching}

\section{Problem statement}\label{sec:problemStatement}
\input{problemStatement}

\section{Proposed solution}\label{sec:solution}
\input{solution}

\subsection{Routing}\label{sec:routing}
\input{solution_routing}

\subsection{Caching}\label{sec:caching}
\input{solution_caching}

\section{Experimentation}\label{sec:experimentation}
\input{experimentation}

\subsection{Testbed setup}
\input{testbedSetup}

\subsection{Network performance}
\input{result_sprintlink}

\section{Conclusion}\label{sec:conclusion}
\input{conclusion}

\appendix[Renewable energy background]\label{sec:renewable}
\input{renewableEnergy}


\ifCLASSOPTIONcaptionsoff
  \newpage
\fi



\bibliographystyle{IEEEtran}
\bibliography{IEEEabrv,transactions_on_computers_rePgbr}
\end{document}

%% file: abstract.tex
\IEEEcompsoctitleabstractindextext{%
\begin{abstract}

The ICT industry today is placed as one of the
major consumers of energy, where recent reports have also
shown that the industry is a major contributor to global carbon
emissions. 
While renewable energy-aware data centers have
been proposed, these solutions have certain limitations. The
primary limitation is due to the design of data centers which
focus on large-size facilities located in selected locations.
This paper
addresses this problem, by utilizing in-network caching with each
router having storage and being powered by renewable energy
sources (wind and solar). Besides placing contents closer to end
users, utilizing in-network caching could potentially increase
probability of capturing renewable energy in diverse geographical
locations. 
Our proposed solution is dual-layered: on the first
layer a distributed gradient-based routing protocol is used to
discover the paths along routers that are powered by the highest
renewable energy, and on the second layer, a caching mechanism
will pull the contents from the data centre and place them
on routers of the paths that are discovered by our routing
protocol. Through our experiments on a testbed utilizing real
meteorological data, our proposed solution has demonstrated
increased quantity of renewable energy consumption, while reducing the workload on the data centers.

\end{abstract}

\begin{keywords}
renewable energy, energy-aware routing, in-network caching
\end{keywords}}

%% file: introduction.tex
\IEEEPARstart{T}{he} current global energy depletion is a widely debated topic, due largely to the increasing population which has fueled massive industrial growth in the last few decades. 
Paralleled to the pressing problem of energy depletion, is the increase in $CO_2$ emissions~\cite{Kocaoglu2012} which has affected the environment in the form of global warming. These problems 
have largely been attributed to the poor planning process and short term goals that we as humans have taken in utilizing natural resources, over the years. 
We are now witnessing a closer relationship between ICT and its influence in the energy sector~\cite{Chiarandini2014,Chung2014}. 
In particular, the growth of the Internet is slowly moving up the ranks as a major source for energy consumption (10\% of the world global energy consumption \cite{Chiaraviglio2012}), which is close to other 
established industries (e.g. airline industry). The wide spread popularity of the Internet has led to an increasing number of deployed communication networks (e.g. WiFi, WiMAX) 
providing rich services (e.g. Multimedia contents) to end users' devices. These rich services are usually represented through contents that are placed in 
high powered data centers~\cite{Zhan2013}, which today is one of the major source of energy consumption in the whole Internet infrastructure.

This new landscape has shifted ICT researchers towards developing solutions that can improve energy consumption of communication networks, and at the same time minimize $CO_2$ 
emissions. In particular, as we witness increasing developments in renewable energy infrastructure, ICT researchers are pursuing new solutions where clean energy could 
be used as an energy source for the Internet infrastructure (e.g. designing energy efficient networks). While new solutions have been proposed for increasing renewable energy 
sources for data centre networks, the limitation of these solutions are the fact that the number of data centers are only localized in small number of locations, which also minimizes 
the probability of meeting high quantity of renewable sources (e.g. as we know, renewable energy is highly dependent on the weather patterns as well as location). However, compressing the size of data 
centers and increasing their distributed locations, will require a brand new design, which will incur high infrastructure costs. 
Therefore, a more feasible and practical solution is required for distributing the source of content storage, that is flexible, highly dynamic and reactive to the changing weather patterns.

In this paper, we propose a solution that meets this objective, by utilizing content routers as a distributed source for content storage. Storing contents on the routers provides a 
number of appealing benefits. Firstly, storing contents within the network provides an opportunity to bring the contents closer to the clients, which minimizes the need to re-fetch 
the contents from end data centers. Secondly, the lower costs of storage prices, means that changes in the infrastructure costs would be minimal in order to enable content caching within 
the network. However, the main benefit that also suits our proposed solution, is the fact that contents can be stored in a distributed manner and increases the distribution of locations 
and access to wider renewable energy sources (we assume that each router is powered by renewable energy infrastructure, which could be a combination of wind turbines and solar panels). 
Therefore, by pulling contents out from the data centers and placing them on content routers which are powered by renewable energy, we will have an opportunity of accessing higher quantity of 
renewable energy to power the content, which in turn can allow us to power-off certain servers in the data centers that may utilize brown energy.

The novelty of our work lies in the ability to maximize the use of
renewable energy to power storage points for contents. Our solution is dual-layered:
\begin{itemize}
\item The first layer uses a gradient-based routing algorithm to discover paths along the routers that have access to high renewable energy.
\item The second layer uses a CCN-like~\cite{jacobson:ccn} in-network caching to cache the contents along the discovered paths.
\end{itemize}

A thorough evaluation has been conducted using real renewable energy data, on a real network topology. 
The evaluation was validated using an experimental testbed. Our key findings can be summarized as follows:

\begin{itemize}
\item The combination of our routing and caching is very effective at
  both increasing use of renewable energy as well as reducing traffic
  in the network (by up to 35\%) even when the renewable infrastructure differ greatly 
  (each geographical location has its own optimal combination of wind turbine or solar 
  panel infrastructure depending on availability of sunlight and wind).
  \item Increased use of renewable energy in the network is equivalent
    to reduction of brown energy; our solution was able to save brown energy usage between 10-$55\%$.
\item  We identify trade-offs between the amount of renewable energy and
  caching performance and show how they interact with each other.
\item Content traffic to data centers
  can be reduced by 24--53\% using the adequate caching strategy which directly translates into energy
  savings at the data center.
\end{itemize}

The paper is organized as follows: Section~\ref{sec:relatedWork} presents the related work.
Section~\ref{sec:problemStatement} clearly defines the objectives of our solution, and this is followed by
Section~\ref{sec:solution} which describes our proposed approach.
Section~\ref{sec:experimentation} describes the results of our experiments.
Section~\ref{sec:conclusion} summarizes the paper, and finally,
the appendix describes the meteorological background used for our study, and the mechanisms of converting renewable energy to consumable power.

%% file: relatedWork_routing.tex
Developing ideas for a greener Internet has been investigated for a
number of years~\cite{Gupta_Singh_2003,Jiang2013}, where proposed solutions
include new routing approaches as well as considering renewable energy
as a possible source of power.

\subsubsection{Energy-aware protocols}

Bolla et al.~\cite{Bolla2011} developed a new approach for next-generation backbone network devices that have smart stand-by primitives through virtualization of the physical network 
infrastructure. 
Their evaluation showed the ability to manage hardware wakeup and standby events transparently from the network-layer protocol. 
Cianfrani et al.~\cite{Cianfrani2010} investigated a simple modification of a link-state routing protocol to minimize the number of links powered-on in a topology.
This led to traffic being redirected to a small number of nodes, yielding reductions in energy consumption. 
Solutions based on energy optimization techniques have also been proposed. In~\cite{Chiaraviglio2012}, an optimization approach was proposed to minimize the traffic and energy consumption of 
the network. 
However, these problems are NP-complete which makes them unrealistic solutions for large-scale network.
An important requirement is to have a more adaptive~\cite{Samaan2009} power-aware routing mechanism that constantly follows the traffic behavior, while reducing the energy consumption without disrupting the QoS requirements of 
the network~\cite{Xiong2011}.

\subsubsection{Renewable energy}

Liu et al.~\cite{Liu_Lin_Wierman_Low_Andrew_2011} proposed the
use of renewable energy for powering data-centers, including an
optimal mix of renewable sources of energy,
using a \kW{30} wind turbine and a \kW{4} solar panel. 
The authors found the optimal energy proportion to be 80\% wind and 20\% solar, which is mainly due to the extra power than can be generated by a \kW{30} wind turbine compared to a 
\kW{4} solar panel.
Unlike~\cite{Liu_Lin_Wierman_Low_Andrew_2011}, we do not focus on a
single data center; instead our goal is to maximize the use of
renewable energy to power the routers within the network.

%% file: relatedWork_caching.tex
Information-centric networking (ICN)~\cite{jacobson:ccn,psirp,Ahlgren:2008:DCN:1544012.1544078,koponen:dona}
has emerged as a general, network-wide caching solution. In ICN the
contents are cached in the network (e.g., in routers), where they are able to serve requests that 
pass through the routers. 

A number of research works have investigated and analyzed the performance of ICNs. In \cite{carofiglio:itc11}, Carofiglio et al. provided an analytical
model for data transfers in ICN.  Muscariello et
al.~\cite{muscariello:icn11} analyzed the ICN performance by taking
into account different bandwidth and storage limits. The solution proposed by  \cite{6193511}
evaluated the ICN performance with realistic traffic mix, showing
there is significant difference between different traffic types
(Web, file sharing, UGC, VoD, etc.). In all these solutions, only
simple Least Recently Used (LRU) approach was used as a caching strategy. Age and popularity-based methods
have also been investigated, where solutions have been developed to placing contents closer to the edge caches~\cite{6193504, 6193512}. In \cite{6193506}, 
an investigation was conducted on the
allocation of cache sizes across the ICN, where results from the analysis showed that the
node's cache size should be proportional to its degree.
In our previous work~\cite{WongW:Caching}, we investigated effects of
cache admission policies and cooperative caching and found them to be
crucial towards good caching performance. The variant of ICN that we
consider in this paper is based on~\cite{WongW:Caching}, which bears a
close resemblance to CCN/NDN~\cite{jacobson:ccn}, in that any router
on the path of a request may answer it, if it has the content cached.

Our work in this paper differs markedly from the work above. Our focus
is not on optimizing caching performance, but instead we combine
caching with renewable energy-aware routing and evaluate the combined solution in
terms of its ability to exploit renewable energy available at
different routers in the network, as well as having a positive influence on the data centers' energy savings.

%% file: problemStatement.tex
\begin{figure}[!t]
\centering
\includegraphics[width=1\figurewidth]{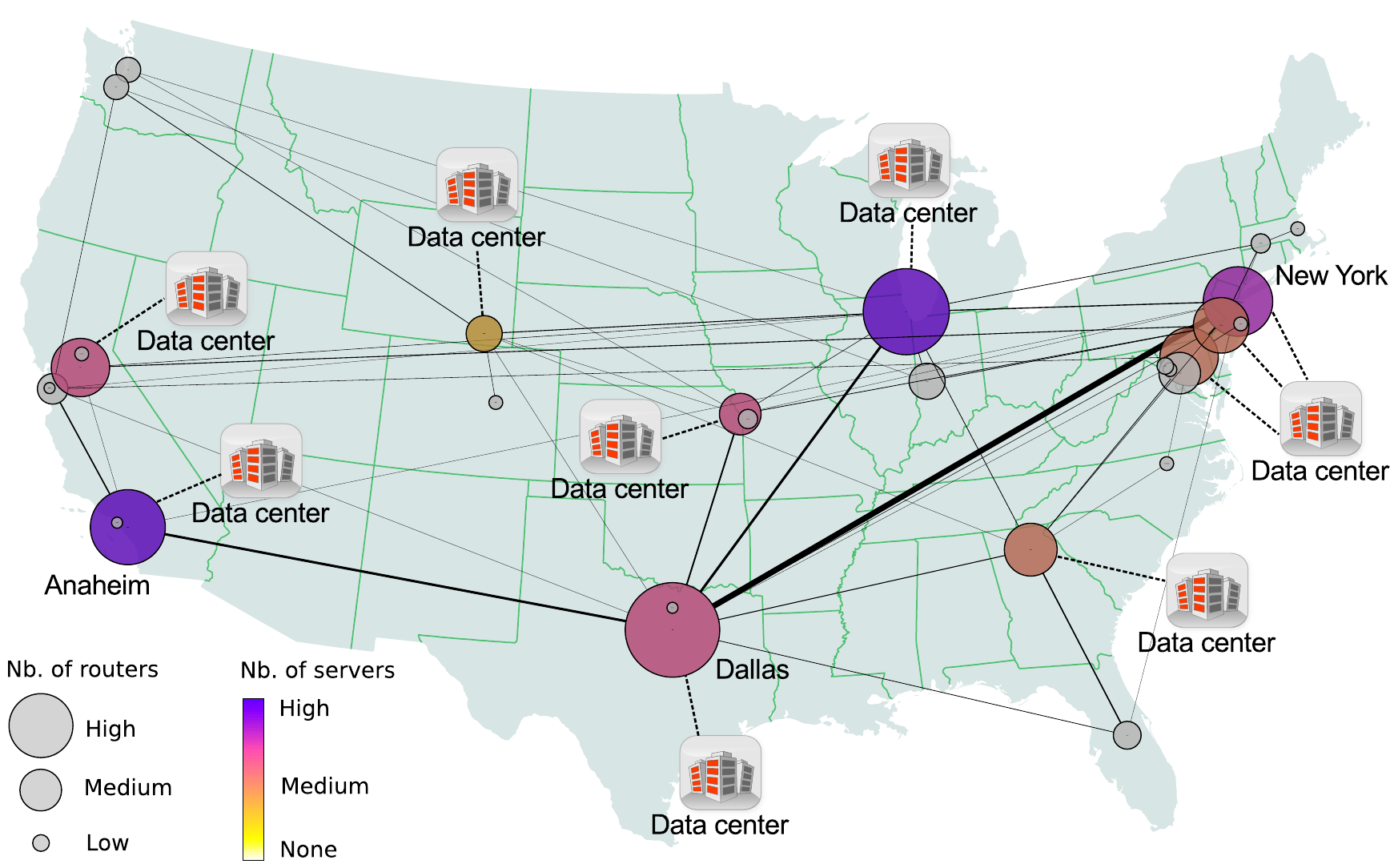}
\caption{Sprintlink USA mainland network.
The figure shows the network links and the size of the bubbles
indicates the size of the POP.
The placement of data centers is explained in Section~\ref{sec:experimentation}.}
\label{fig:spintlinkMainland}
\end{figure}

We now define our problem statement, and models that we use to represent the renewable energy consumption by the routers of the network. We consider an Internet Service Provider (ISP) network (such as \figurename{ \ref{fig:spintlinkMainland}}), 
in which nodes may be powered partially from renewable energy sources, where this source could be a combination of wind and solar. The ISP network connects the data centers and the users access points, 
which serve as the source and destination points for traffic in the network. The remaining routers in the ISP network serve as transport nodes that 
form the ISP's mesh topology. 

An informal description of the design problem we consider is the following:
\begin{description}[\setlabelwidth{\vspace{-50em}}]
 \item{[\textbf{Given}]} (i) a realistic ISP network topology composed of routers and bi-directional links, (ii) an infrastructure for renewable energy for each router, 
(iii) the power consumption of all devices in the networks besides data centers, and finally (iv) a model for content popularity in function of time.
 \item{[\textbf{Objective}]} is to (i) determine the impact of integrating renewable energy awareness on the network and the data center performance, and (ii) if needed, determine the best trade-off
 to maximize the reduction of brown energy for the ISP networks and
 the data centers, while maintaining satisfactory content delivery performance for the end users.
\end{description}

The energy consumption model of the ISP's network devices is derived from the study of Chabarek et al. \cite{Chabarek_Sommers_Barford_Estan_Tsiang_Wright_2008}, which was 
extended from the model proposed initially by Gupta and Singh \cite{Gupta_Singh_2003}.  
The model proposed by \cite{Gupta_Singh_2003} was developed by empirically monitoring the energy consumption of chassis and line-cards in two Cisco routers (\emph{Cisco GSR 12008 and 7507}) under various traffic loads. 
Results from the experiments showed that the chassis is the biggest consumer of energy while the line-cards consume less. 

In this paper, a new energy consumption model for routers has been developed to incorporate use of renewable energy.
\cite{Chabarek_Sommers_Barford_Estan_Tsiang_Wright_2008}. The general energy model for a router's power consumption is represented as:
\begin{equation}
  PC(X) = CC(X_0) + \displaystyle\sum_{i=1}^N (TP(X_{i}) + LCC(X_{i}))%
  \label{eq:original_chabarek}
\end{equation}
where $X$ is the vector of chassis and line-cards energy models, as well as the traffic configuration of the router. $CC(X_0)$ is the energy consumption for the chassis, 
$N$ is the number of line-cards of router $X$, 
$TP(X_i)$ is the energy consumption due to traffic on the line-card $i$, and $LCC(X_{i})$ is the line-card energy consumption.

Since the impact of traffic load on the routers does not fluctuate significantly, the $TP(X_i)$ can be omitted in equation \ref{eq:original_chabarek}, leading to:
\begin{equation}
  PC(X) = CC(X_0) + \displaystyle\sum_{i=1}^N LCC(X_{i})
  \label{eq:reduced_chabarek}
\end{equation}

In our energy consumption model, we assume that a router has the ability to power-off line-cards, and the entire router will be powered-off when all its line-cards are off. 
Therefore, the power consumption model at time $t$ can be represented as follows:
\begin{equation}
\label{eq:router_energy_model}
  PC(X,t) = x_{0,t} CC(X_0) + \displaystyle\sum_{i=1}^N x_{i,t} LCC(X_{i})
\end{equation}
where
\begin{align*}
  x_{0,t}&=\begin{cases}
  0& \text{if $\sum_{i=1}^N (x_{i,t}) = 0$}\\
  1& \text{otherwise}
  \end{cases}\\
  x_{i,t},i > 0&=\begin{cases}
  0& \text{if $X_i$ is powered-off at time $t$}\\
  1& \text{if $X_i$ is powered-on at time $t$}
  \end{cases}
\end{align*}

Consequently, the configuration of each $x_{i,t}$ is added to the vector $X$. Let $X^0$ be the vector of chassis and line-cards models, where all $x_{i,t} = x_{i,t}^0 = 1$.

Let us assume that each router $X$ would have a source of renewable energy $rePC(X,t)$ at time $t$:
\begin{equation}
  rePC(X,t) = P_w(X,t) + P_s(X,t)
\end{equation}
where $P_w(X,t)$ and $P_s(X,t)$ are respectively the power generated from wind and solar energy at time $t$. More details can be found in the appendix on the methods used to calculate $rePC(X,t)$ using real meteorological data.

By subtracting $rePC(X,t)$ from $PC(x)$, we obtain $brPC(X,t)$, the brown energy consumed by router $X$ at time $t$.
\begin{multline}
  brPC(X,t) = PC(X^0,t) - rePC(X,t)\\\text{with $brPC(X,t) = 0$, if $rePC(X,t) \geq  PC(X^0,t)$}
\end{multline}

Our aim in this paper is to analyse the impacts of favoring routers powered by renewable energy in order to facilitate the powering-off of the unused routers that may be powered by brown energy.
Therefore, we can formulate the brown energy reduction of the entire network $\sigma_{n}$ at time $t$ by calculating the total brown energy consumption of the network when unused devices are powered-on or off. 

\begin{equation}
  \sigma_{n} = 1 - \frac{\displaystyle\sum_{\forall X}\displaystyle\sum_{\forall t} brPC(X,t)}{\displaystyle\sum_{\forall X}\displaystyle\sum_{\forall t} brPC(X^0,t)}
  \label{eq:brownSavingsNetwork}
\end{equation}
 
Let $\alpha$, $\alpha \in [0,1]$ be the factor that favor renewable energy, and where $\sigma_{n}(\alpha)$ and $\sigma_{d}(\alpha)$ are respectively the brown energy 
reduction utility functions of the network and the data centers derived from the choice of $\alpha$. 
In the subsequent sections we will show how $\alpha$ is used in our renewable energy aware gradient-based routing algorithm.
While our paper does not focus on the utility function of the data centers, we discuss the trade-off that can be obtained by changing $\alpha$ and 
show that the $\alpha$ value can be optimized according to seasonal changes. 

To conclude, we want to maximize $\sigma_{n}(\alpha) +
\sigma_{d}(\alpha)$, while minimizing the impact on the performance for the end users.

%% file: solution.tex
\begin{figure}[!t]
\centering
\includegraphics[width=\figurewidth]{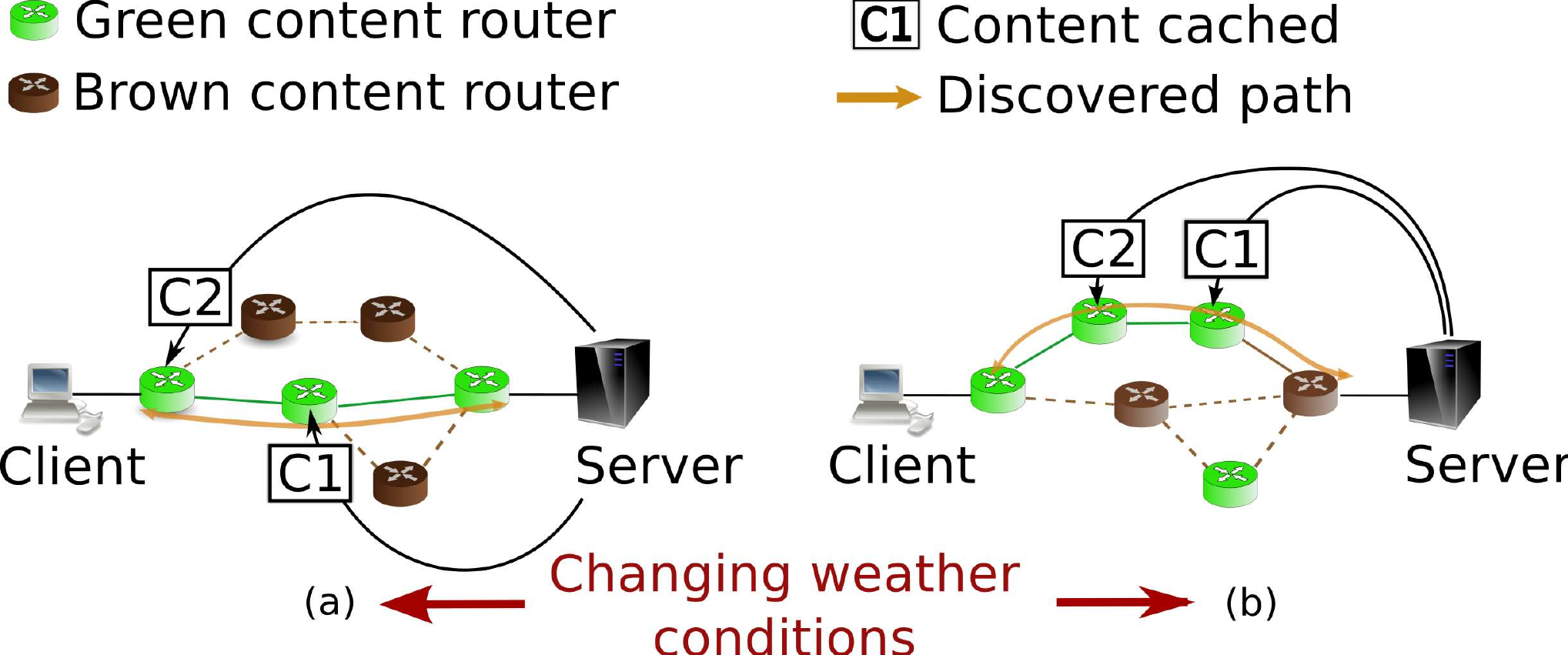}
\caption{Illustration of proposed approach, demonstrating path discovery as weather condition changes ((a) to (b)). Once paths are discovered, contents C1 and C2 
are cached on routers along the discovered paths.}
\label{fig:proposedSolution}
\end{figure}

Fig.~\ref{fig:proposedSolution} shows an example of our proposed approach, where routes are discovered via routers with access to high renewable energy. 
Once the paths are discovered, contents will be pulled from the data centre and populated along the discovered paths. 
However, these paths will change as the weather pattern changes, which implicitly moves the contents to new locations that have access to high renewable energy. 
At the same time, certain routers that are powered by ``brown'' energy maybe unused, which could lead to a situation where we can power-off these routers to minimize overall brown energy.
Therefore, as shown in Fig.~\ref{fig:proposedSolution}, the proposed approach is dual-layered.

Firstly, a gradient-based routing protocol discovers the path of routers powered by 
the highest amount of renewable energy, and secondly a CCN-like~\cite{jacobson:ccn} caching approach populates the routers of the discovered path with
contents based on a caching strategy. Therefore, the approach maximizes the dynamic properties of caching, where contents can be moved to different locations, and in 
this particular case to locations that provide the highest amount of green energy. 

\begin{figure}[!t]
  \centering
  \subfloat[Wind speed]{\includegraphics[width=0.5\figurewidth]{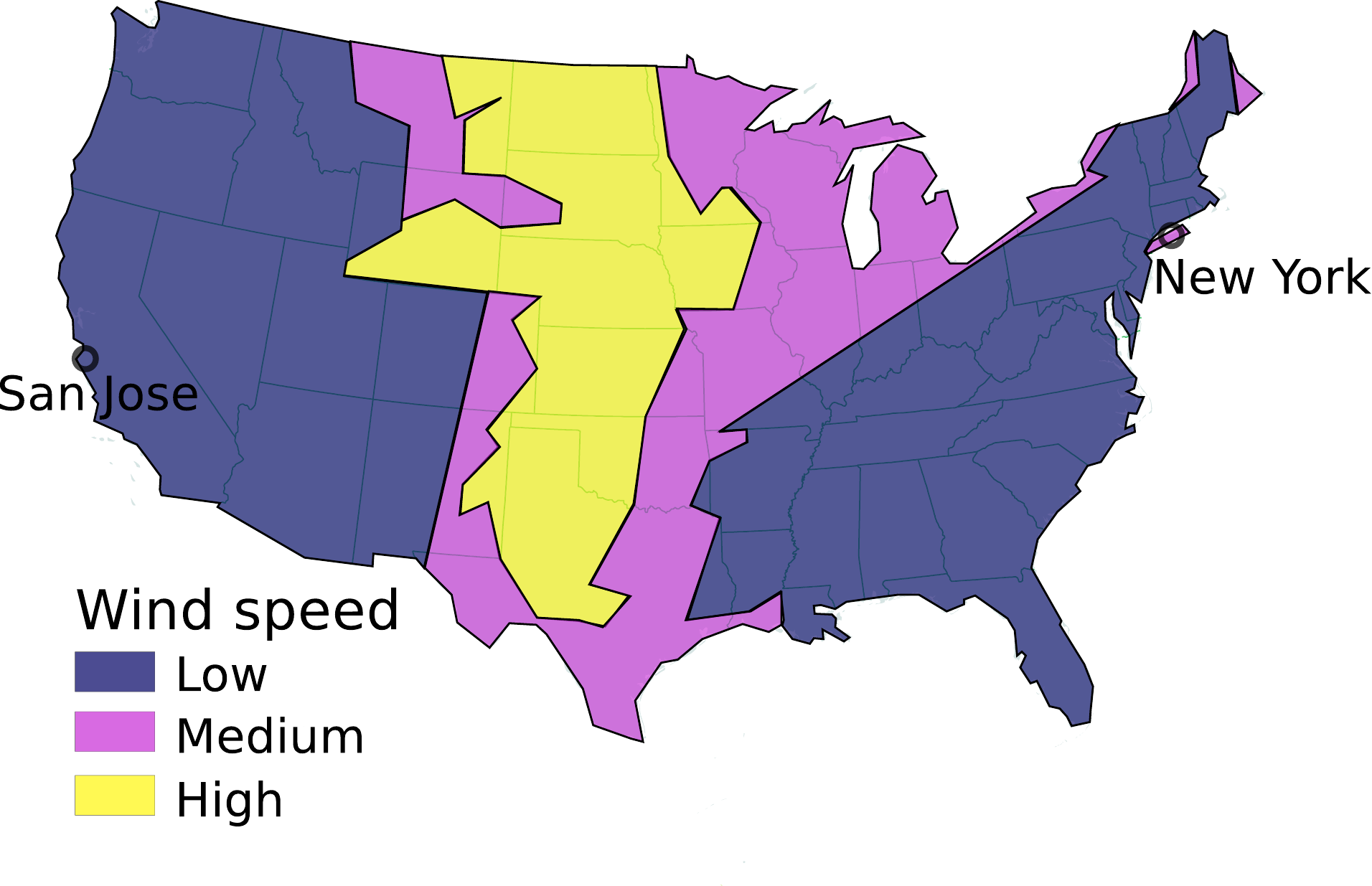}%
    \label{fig:windSpeed}}
  \subfloat[GHI]{\includegraphics[width=0.5\figurewidth]{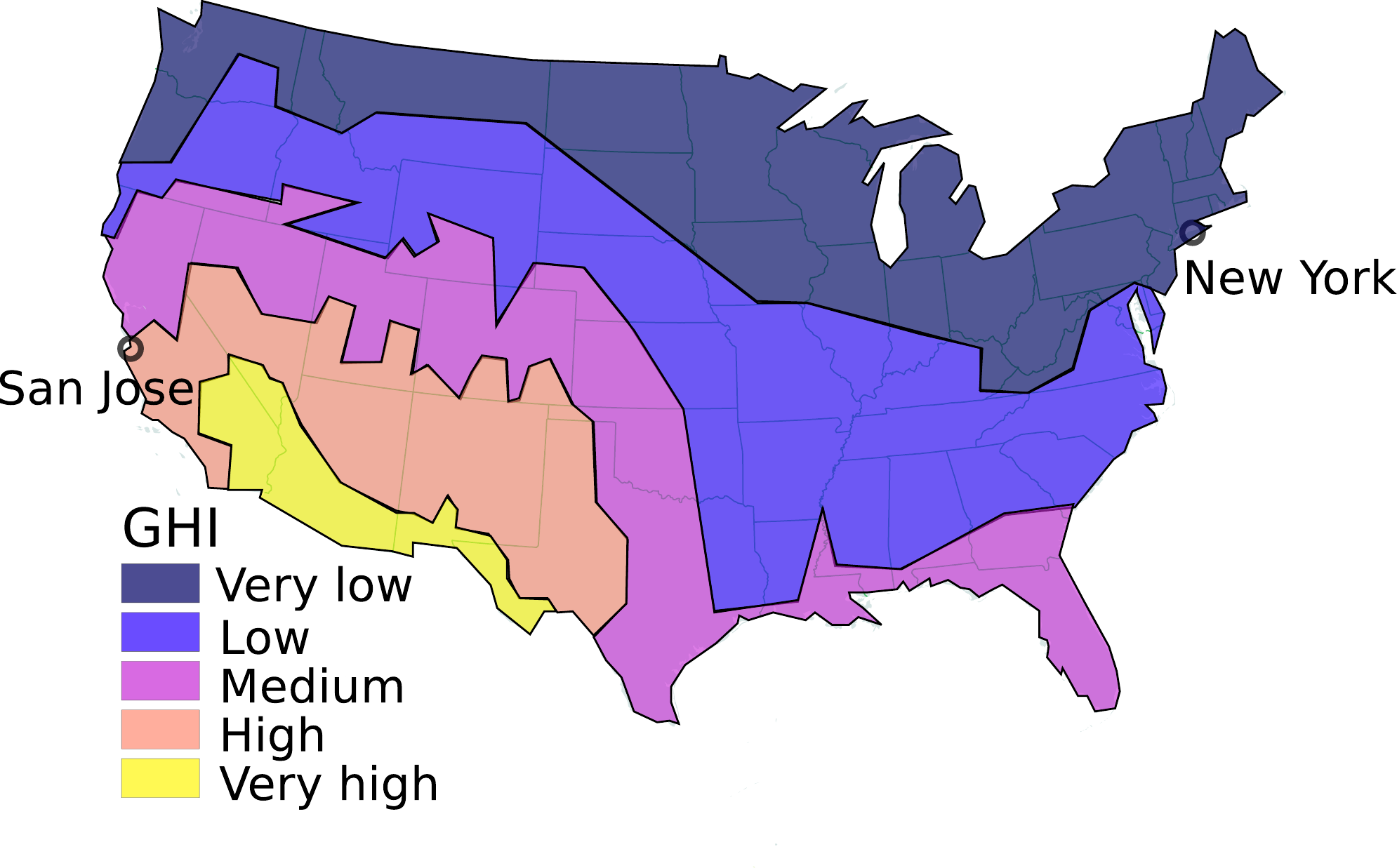}%
    \label{fig:ghi}}
  \caption{Average annual wind speed and Global Horizontal Irradiance in USA.
  Brighter colors indicate higher availability.}
  \label{fig:usaWeather}
\end{figure}

An important factor that affects the performance of renewable energy production is the location. For instance, \figurename{~\ref{fig:usaWeather}} shows the average annual wind speed 
and Global Horizontal Irradiance in the USA. 
As shown by these two figures, the available energy sources greatly differ depending on the locations of routers. Cities such as New York or San Jose exhibits 
dissimilar weather conditions, including
predictability in the weather patterns.  
More detailed information can be found in the appendix which discusses the quantity of wind and solar energy for specific cities. 
While the production of renewable energy fluctuates for different locations, this is most ideal for our gradient-based routing protocol. 
The gradient-based routing protocol, which is highly scalable and distributed, can discover routes by
adapting to these weather conditions in order to maximize the renewable energy. 

The metric used to determine the greenest router is based on the highest ratio of renewable energy consumed compared to total energy requirements of the router $g(X,t)$ (this is also termed \emph{green ratio}). 
This ratio is calculated according to equation \ref{eq:green_ratio}:
\begin{equation}\label{eq:green_ratio}
 g(X,t)=\begin{cases}
  1& \text{if $rePC(X,t) \geq PC(X,t)$}\\
  \frac{rePC(X,t)}{PC(X,t)}& \text{otherwise}
  \end{cases}\\
\end{equation}

Examples of the variations of the green ratio for different locations
are presented in \figurename{~\ref{fig:powerAndData2Days_sanJose_5kW_greenRatio}-\ref{fig:powerAndData2Days_newYork_30kW_greenRatio}}. 
Every router in the network locally broadcasts messages to its one-hop neighbors regarding the ratio of renewable energy that it
has available. The greenest routers, will in turn maximize the production of packets processed with renewable energy (\emph{green packets}). 

The subsequent sections, we describe our routing protocol and caching strategies in more detail.

%% file: solution_routing.tex
Our routing protocol, called \emph{Renewable energy-aware Parameterized Gradient Based Routing} (\rePGBR{}), 
is extended from our original \emph{Parameterized Gradient Based Routing (PGBR)} protocol
\cite{Balasubramaniam2011a}. The original PGBR is a fully distributed bio-inspired routing protocol, that is inspired by the 
\textit{Bacterial Chemotaxis} process. Through the process  of \textit{Chemotaxis}, bacteria are able to mobilize themselves towards a destination point by sniffing a chemical gradient emitted  from the destination node. 
This same bacteria motility principle is used for the \rePGBR{} routing algorithm, where the routes are discovered
by hopping from node to node along the path with the highest gradient, 
until it reaches the destination. The benefit of the \rePGBR{} algorithm, includes (i) high scalability, (ii) no 
requirements for pre-knowledge of the traffic demand, 
and (iii) the ability to efficiently discover paths to contents favoring the usage of renewable energy.
 
The gradient field equation is represented as follows:
\begin{equation}
G_i^d(j) = \alpha g(j) + (1 - \alpha) h_i^{d}(j), \quad 0 \leq \alpha \leq 1
\label{eq:gradient}
\end{equation}
where $G_i^d(j)$ represents the gradient value of the link $i \rightarrow j$ for a packet to destination $d$, $g(j)$ (e.g. equation \ref{eq:green_ratio}) represents the green ratio value of 
neighbor node $j$, and $h_i^d(j)$ represents the normalized hop count value of neighbor node $j$ of node $i$ for destination $d$. 
The $\alpha$, as specified in Section~\ref{sec:problemStatement}, represents the weighting parameter between the shortest path and the greenest path to the content location, 
and is a key parameter in evaluating the trade-off between 
maximizing green energy and maximizing caching performance.  Unlike the normalized hop count value used  in \cite{Balasubramaniam2011a}, 
where the equation for $h_i^d(j)$ is $h_i^d(j) = 1 - \frac{w^d(j)}{W^d}$, with $W^d = max(w^d(k)), \forall k$, 
the new normalized hop count value of \rePGBR{} provides fast destination discovery in a scalable manner \cite{Mineraud2012}. The modified $h_i^d(j)$ is represented as:
\begin{equation}
 h_i^d(j) = \frac{max(w_i^d(k)) - w_i^d(j)}{max(w_i^d(k)) - min(w_i^d(k))}, \forall k \text{ neighbours of } i
\end{equation}
where $w_i^d(k)$ represents the weight of node $k$ in the shortest path tree for root $d$, the destination.

\begin{algorithm}[!t]
\caption{Selection of next hop during discovery}\label{alg:nextHop}
\begin{algorithmic}[1]
\REQUIRE Destination $d$, incoming neighbour $iN$
\STATE $u \leftarrow$ current node
\STATE $qIn \leftarrow$ list of incoming neighbours
\STATE $qOut \leftarrow$ list of outgoing neighbours

\IF{$iN$ is not nil}
  \STATE $qIn \leftarrow qIn + iN$
\ENDIF
 
\STATE $nextHop \leftarrow nil$
\STATE $bestG \leftarrow -1$

\FOR{neighbor $v$ of $u$}
  \IF{$v \notin qIn \cup qOut \text{ and } G_u^d(v) > bestG$}\label{alg:line_green}
    \STATE $bestG \leftarrow G_u^d(v)$
    \STATE $nextHop \leftarrow v$
  \ENDIF
\ENDFOR

\IF{$nextHop$ is nil}\label{alg:backtracking_start}
  \REPEAT
  \STATE $nextHop \leftarrow last(qIn)$
  \UNTIL{$nextHop \notin qOut$}
\ENDIF\label{alg:backtracking_end}
 
\IF{$nextHop$ is not nil}
  \STATE $qOut \leftarrow qOut + nextHop$
\ENDIF

\RETURN $nextHop$
\end{algorithmic}
\end{algorithm}

\input{discovery1}

Routing is performed by first discovering the path between a source-destination pair, by sending a discovery packet that migrates hop-by-hop, selecting the link with the highest
gradient value. Algorithm \ref{alg:nextHop} describes how to choose the next hop during discovery. Unlike the original PGBR where the path was stored in the discovery packet,
\rePGBR{} stores local information in routers. This local information enables the discovery to avoid loops 
(i.e. using list of incoming and outgoing neighbours, line \ref{alg:line_green}) and performs back-tracking 
(i.e. lines \ref{alg:backtracking_start} to \ref{alg:backtracking_end}) in a fully and efficient distributed manner.

Furthermore, this discovery process is depicted in \figurename{~\ref{fig:route_disc1}} and \figurename{~\ref{fig:route_disc2}} illustrates the discovery process dynamically changing to avoid 
brown hot-spot areas (e.g. discovery avoids brown node 5 due to the gradient field changes). Once the weather conditions change, a new discovery is issued. However, during the discovery process, 
the existing path remains stable until the new discovered path is adopted sequentially as the discovery message backtracks to the source using the local information described in Algorithm 
\ref{alg:nextHop}.

The utilization of the weighting parameter $\alpha$ provides \rePGBR{} with robustness in supporting multiple objectives. 
For instance, when $\alpha$ is set to a low value (e.g. $0.2$), the 
discovered path is biased towards the shortest routes (more weight is attributed to the hop count) while a larger value would 
tend to favor greener routes. As a result, the gradient field 
equation of \rePGBR{} is flexible and can exhibit multiple behaviors by only manipulating one parameter to provide the desired green performance of 
the network. This paper aims to investigate the most optimum $\alpha$, that could maximize the savings of brown energy usage of the full network, as well as
having a positive influence on the data centers' energy savings.

%% file: discovery1.tex
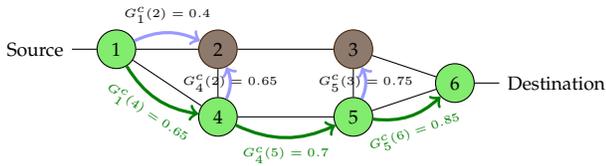
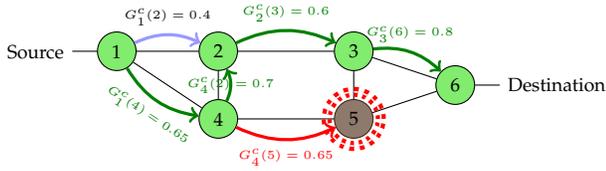
\begin{figure}[t]
\centering
\definecolor{greenPgbr}{rgb}{0.1765,0.878,0.047}
\definecolor{brownPgbr}{rgb}{0.255,0.129,0.047}
\subfloat[Discovery of paths along gradient field formed by routers with high renewable energy.]{
\label{fig:route_disc1}
\begin{tikzpicture}[scale=0.9,transform shape]
 	\node	(S) at (0.3,1) {\footnotesize Source};
	\node	(1)	at (1.5,1)[shape=circle,draw=black,fill=greenPgbr!60] {\footnotesize 1}
		edge (S);
	\node	(2)	at (3,1)[shape=circle,draw=brownPgbr,fill=brownPgbr!60] {\footnotesize 2}
		edge (1)
		edge[<-,style=very thick,draw=blue!40,bend right=25] node[above] {\tiny $G_1^c(2)=0.4$} (1);
	\node	(3)	at (5,1)[shape=circle,draw=brownPgbr,fill=brownPgbr!60] {\footnotesize 3}
		edge (2);
	\node	(6)	at (6.5,0.5)[shape=circle,draw=black,fill=greenPgbr!60] {\footnotesize 6}
		edge (3);
	\node	(D)	at (8,0.5) {\footnotesize Destination}
		edge (6);
	\node	(5)	at (5,0)[shape=circle,draw=black,fill=greenPgbr!60] {\footnotesize 5}
		edge (6)
		edge (3)
		edge[->,style=very thick,draw=blue!40,bend right=25] node[midway] {\tiny $G_5^c(3)=0.75$} (3);
	\node	(4)	at (3,0)[shape=circle,draw=black,fill=greenPgbr!60] {\footnotesize 4}
		edge (5)
		edge (1)
		edge (2);
	\draw [<-,style=very thick,draw=green!50!black!100] (4)	to [bend left=25] node[midway,sloped,below]
				{\textcolor{green!50!black!100}{\tiny $G_1^c(4)=0.65$}} (1);
	\draw [->,style=very thick,draw=green!50!black!100] (4)	to [bend right=25] node[midway,below]
			{\textcolor{green!50!black!100}{\tiny $G_4^c(5)=0.7$}} (5);
	\draw [->,style=very thick,draw=green!50!black!100] (5)	to [bend right=25] node[midway,below,sloped]
				{\textcolor{green!50!black!100}{\tiny $G_5^c(6)=0.85$}} (6);
	\draw [->,style=very thick,draw=blue!40] (4)	to [bend right=25] node[midway]
				{\tiny $G_4^c(2)=0.65$} (2);
	\end{tikzpicture}
}\\
\subfloat[Gradient field modification due to changes in renewable energy performance.]{
\label{fig:route_disc2}
\begin{tikzpicture}[scale=0.9,transform shape]
 	\node	(S) at (0.3,1) {\footnotesize Source};
	\node	(1)	at (1.5,1)[shape=circle,draw=black,fill=greenPgbr!60] {\footnotesize 1}
		edge (S);
	\node	(2)	at (3,1)[shape=circle,draw=black,fill=greenPgbr!60] {\footnotesize 2}
		edge (1)
		edge[<-,style=very thick,draw=blue!40,bend right=25] node[above] {\tiny $G_1^c(2)=0.4$} (1);
	\node	(3)	at (5,1)[shape=circle,draw=black,fill=greenPgbr!60] {\footnotesize 3}
		edge (2);
	\node	(6)	at (6.5,0.5)[shape=circle,draw=black,fill=greenPgbr!60] {\footnotesize 6}
		edge (3);
	\node	(D)	at (8,0.5){\footnotesize Destination}
		edge (6);
	\node	(5)	at (5,0)[shape=circle,draw=brownPgbr,fill=brownPgbr!60] {\footnotesize 5}
		edge (6)
		edge (3);
	\node	(4)	at (3,0)[shape=circle,draw=black,fill=greenPgbr!60] {\footnotesize 4}
		edge (5)
		edge (1)
		edge (2);
	\path [draw=red,double,dotted,ultra thick] (5,0) circle (0.4);
	\draw [<-,style=very thick,draw=green!50!black!100] (4)	to [bend left=25] node[midway,sloped,below]
				{\textcolor{green!50!black!100}{\tiny $G_1^c(4)=0.65$}} (1);
	\draw [->,style=very thick,draw=red] (4)	to [bend right=25] node[midway,below]
			{\textcolor{red}{\tiny $G_4^c(5)=0.65$}} (5);
	\draw [->,style=very thick,draw=green!50!black!100] (2)	to [bend left=25] node[midway,above]
		{\textcolor{green!50!black!100}{\tiny $G_2^c(3)=0.6$}} (3);
	\draw [->,style=very thick,draw=green!50!black!100] (4)	to [bend right=25] node[midway]
				{\textcolor{green!50!black!100}{\tiny $G_4^c(2)=0.7$}} (2);
	\draw [->,style=very thick,draw=green!50!black!100] (3)	to [bend left=25] node[midway,above]
		{\textcolor{green!50!black!100}{\tiny $G_3^c(6)=0.8$}} (6);
	\end{tikzpicture}
}
\label{fig:route_discovery}
\caption{Route discovery process for \rePGBR{}.}
\end{figure}

%% file: solution_caching.tex
This section will describe the second component of our solution, which is the caching process of contents along the paths discovered by \rePGBR{}. We used three caching strategies described in~\cite{WongW:Caching}. 
The work in~\cite{WongW:Caching} focused on evaluating these strategies purely on caching performance across a 
wide range of networks. In this paper, we focus on using these caching strategies to maximize the use of renewable energy for specific points of storage, and the suitability of these strategies in combination with the 
adapted \rePGBR{} routing algorithm. Below we describe the three strategies and briefly recap the main results from~\cite{WongW:Caching}.
In terms of different information-centric architectures, our caching strategies would be most suitable for a CCN/NDN-like architecture~\cite{jacobson:ccn}.

{\bf ALL}: ALL is the simplest caching strategy, where each router tries to cache all the packets that passes through, and uses LRU algorithm to evict packets if the 
cache is full. There is no cooperation between the routers.

{\bf Cachedbit}: Cachedbit uses one bit in the packet header to indicate whether the packet has already been cached by a router along the path. If the bit is not set, a router decides 
locally whether to cache the packet or not, with a probability of $1/n$ where $n$ is the length of the path from the client to the server (or ingress router to egress router for an intermediate network). 
As discussed in~\cite{WongW:Caching}, the length of the path can be determined by observing packets that pass through a router; global knowledge is not required. In this case, the caches use LRU as a 
replacement policy for evicting the contents. Algorithm~\ref{alg:cachedbit} describes the admission policy for the Cachedbit strategy.

\begin{algorithm}[!t]
\caption{Admission Policy in Cachedbit Strategy}
\label{alg:cachedbit}
\begin{algorithmic}[1]
  \STATE{\textbf{Input:} Data chunk $C_{i}$}
  \STATE{\textbf{Output:} Caching decision}
  \IF {$C_{i}$ is not cached \AND $cached\_bit$ not set}
  \STATE {Draw a random $x$ uniformly from $(0,1)$}
  \IF {$x < 1/n$}
  \IF {Cache is full}
  \STATE {Evict entry based on LRU}
  \ENDIF
  \STATE {Add data chunk $C_{i}$}
  \STATE {Set $cached\_bit$}
  \ENDIF
  \ENDIF
  \STATE {Forward chunk to destination}
\end{algorithmic}
\end{algorithm}

{\bf Neighbor Search (NbSC)}: 
NbSC is otherwise similar to Cachedbit, except that routers periodically send Bloom filters of their contents to their neighbor routers. When a router experiences a miss, it can check if any 
of its neighbors has the requested packet. If there are multiple matches, the request will be redirected to a random neighbor. We also investigated a variant which redirects requests to the 
greenest neighbor.

The results in~\cite{WongW:Caching} showed that a Cachedbit-like
admission policy is needed to get good caching performance, but that
the addition of NbSC reduces network traffic considerably.

Algorithm~\ref{alg:nbsc} describes the cooperation policy in the algorithm that implements NbSC.

\begin{algorithm}[!t]
\caption{Cooperation Policy in NbSC Strategy}
\label{alg:nbsc}
\begin{algorithmic}[1]
 \STATE{\textbf{Input:} Data request $R_{i}$}
 \STATE{\textbf{Output:} Response decision}
 \IF {$C_i$ is cached for $R_i$}
  \STATE {Reply with $C_i$}
 \ENDIF
 \IF {$C_i$ is not cached for $R_i$}
  \IF {Neighbor $N_i$ cached $C_i$ for $R_i$}
    \STATE {Redirect $R_i$ to $N_i$}
  \ELSE
    \STATE {Forward $R_i$ to next hop}
  \ENDIF
 \ENDIF

\end{algorithmic}
\end{algorithm}

%% file: experimentation.tex
In order to validate our proposed approach, we have conducted
experiments on a testbed using the weather data described
in the appendix. The network evaluated on our testbed is based
on the Sprint router-level topology from the Rocketfuel
project~\cite{Spring2002}. The Sprint network consists of 278 routers
geographically distributed in 27 cities in U.S. mainland (meteorological data from ~\cite{nrel} is available only for the U.S. mainland locations). For the
experiment, the top 40 highest degree nodes are connected to content
servers, while the 80 routers with the lowest degrees are connected to
the clients. \figurename{~\ref{fig:spintlinkMainland}} shows the resulting topology where routers located in the same city are grouped (i.e. the size of the city increases with the number of routers).
For instance, Anaheim, Chicago, New York and Dallas are the cities with the highest number of routers. The color of the cities, on the other hand, represents the number of servers (e.g. nodes which have the highest
connectivity). A total of 10 cities have servers resulting in 10 data centers that are interconnected through the ISP network.
Once again, Anaheim, Chicago and New York are the three cities with the highest number of servers. 

The remaining routers are used as intermediate nodes for the routing algorithm. In our experimental setup,
the clients will continually request data from the servers (request rate depends on the traffic pattern). 
While tests have been conducted for three different traffic patterns, which includes (i) real trace
from the TOTEM project \cite{totem}, (ii) trace from the gravity model
\cite{anand:sigcom09}, and (iii) constant traffic pattern, 
there was no significant difference between these traffic patterns in
terms of performance. Therefore, we only show the results of our
experiments based on a constant traffic pattern.

The request trace for each client is generated from a real DNS trace of
a university lab. We sorted the requested DNS names according to their
popularity and removed the top 20 entries. The request pattern from
this trace follows a Zipf distribution with parameter $0.9$, which is
very close to real-life distribution shown in~\cite{anand:traffic}.
Our trace requests chunks of content, which are assumed to be
independent of each other. We assigned each router with a storage
capacity of 4 gigabytes. 
The power requirements of every router has been set using the measurements of a Cisco 7507 \cite{Chabarek_Sommers_Barford_Estan_Tsiang_Wright_2008}: the basic chassis requiring 
210W and each line-card an additional 70W (i.e. a router with 4 line-cards consumes $210 + 4*70 = 490~W$).

We investigated two scenarios: In scenario A, the renewable energy infrastructure of the routers are set to supply twice the energy required by the router 
(i.e. $rePC(X,t) = 2 PC(X^0,t)$) at most. Therefore, the changes of weather conditions greatly modify the green ratio $g(X,t)$. This scenario is defined as the 
static capacities scenario. In scenario B, the infrastructure supplies up to three times the energy required by the routers. The different size 
of infrastructures implies that some routers are mostly fully brown and others mostly green, which will better represent a realistic scenario where green renewable
energy farms are built gradually (to limit the cost of installation). More details on the settings can be found in the appendix.

The following metrics have been used to evaluate our proposed solution for the two scenarios previously described:

\begin{itemize}

\item \textbf{Hit Rate:} Determines the fraction of requests that are served by
 the content routers. Since all objects are of the same size, the hit rate is also
  the byte hit rate. Hit rate shows how much external traffic is saved
  by caching, thus leading to potential energy savings for the data centers.

\item \textbf{Footprint Reduction:} Network footprint is the product
  of the amount of content and the network distance from which the
  content was retrieved. It measures the amount of internal traffic
  reduction, where a smaller footprint (larger reduction) means less
  traffic within the network.

\item \textbf{Green/brown ratio:} Proportion of the packets by the router within an hour that has been
   processed using renewable energy to the packets that has been
   processed using fossil fuel energy.
 
\item \textbf{Reduction of brown packets:} Reduction of brown packets
   that has been processed within an hour using fossil fuel energy.

\item \textbf{Brown energy savings:} The amount of brown energy that has been saved by powering-off unused routers.
   In the case of a neighbouring router being unused, the line-card connecting the two routers is considered to be powered-off. The brown energy savings is calculated using equation~\ref{eq:brownSavingsNetwork}.

\end{itemize}

While hit rate and the footprint reduction are mainly caching metrics, the green/brown ratio, eduction of brown processed packets and
the brown energy savings are representing the performances of the solution to successfully address the energy efficiency requirements.
Hit rate also represents the reduction on traffic towards data centers and, 
therefore, indicates potential energy savings at the data center.

%% file: testbedSetup.tex
All the experiments are performed on a cluster of 240 Dell PowerEdge M610 nodes. Each node has 2 quad-core CPUs, 32GB memory, and is connected to a 10-Gbit network. 
All the nodes run Ubuntu SMP with 2.6.32 kernel. The experimental platform we used in the evaluation is capable of simulating realistic routers, and allocating necessary 
physical resources according to the simulated network size. In the event that the network size is larger than the actual number of nodes in the cluster, multiple routers will 
be multiplexed onto one node.

%% file: result_sprintlink.tex
\begin{figure*}[!t]
 \centering
 \subfloat[ALL]{\label{fig:footprintStaticAll}\includegraphics[width=0.3\linewidth]{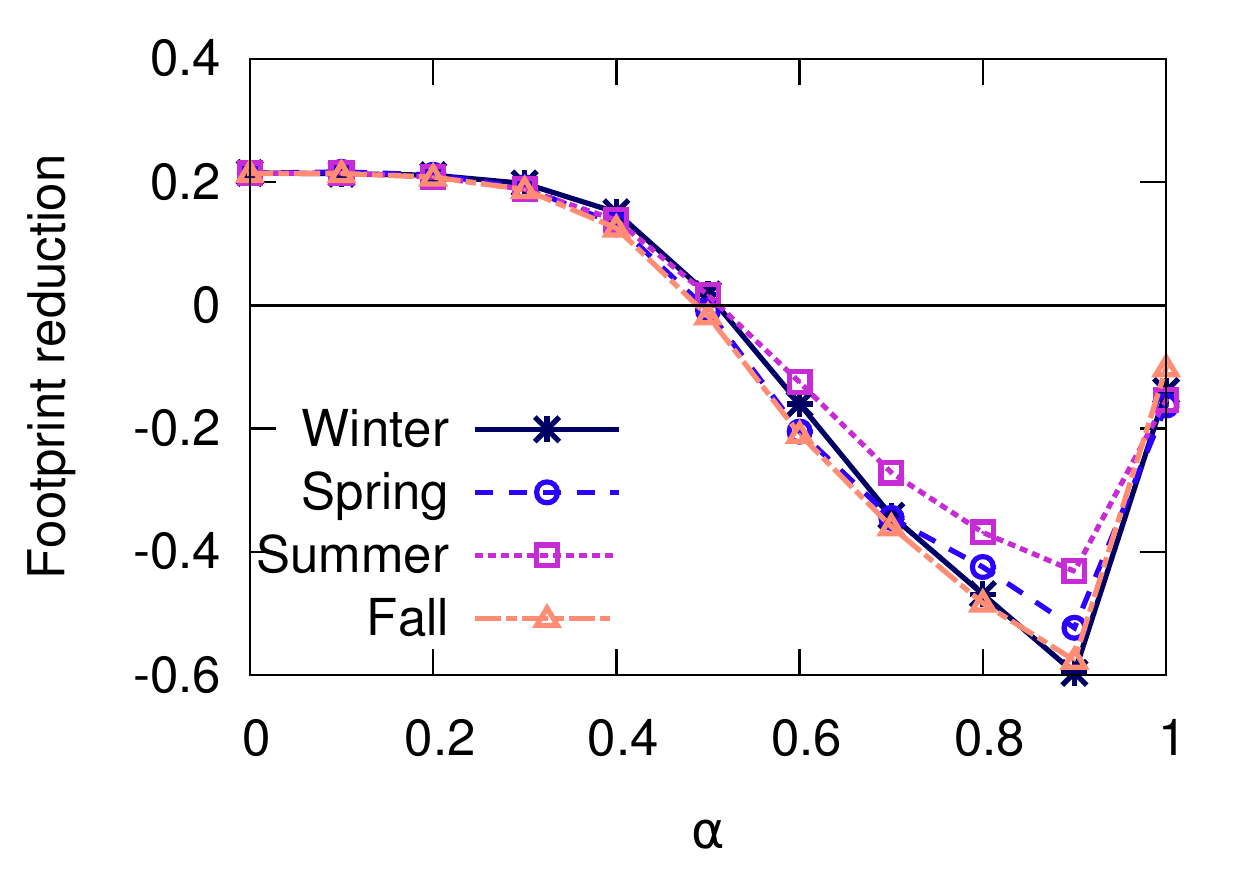}}
 \subfloat[Cachedbit]{\label{fig:footprintStaticCachedbit}\includegraphics[width=0.3\linewidth]{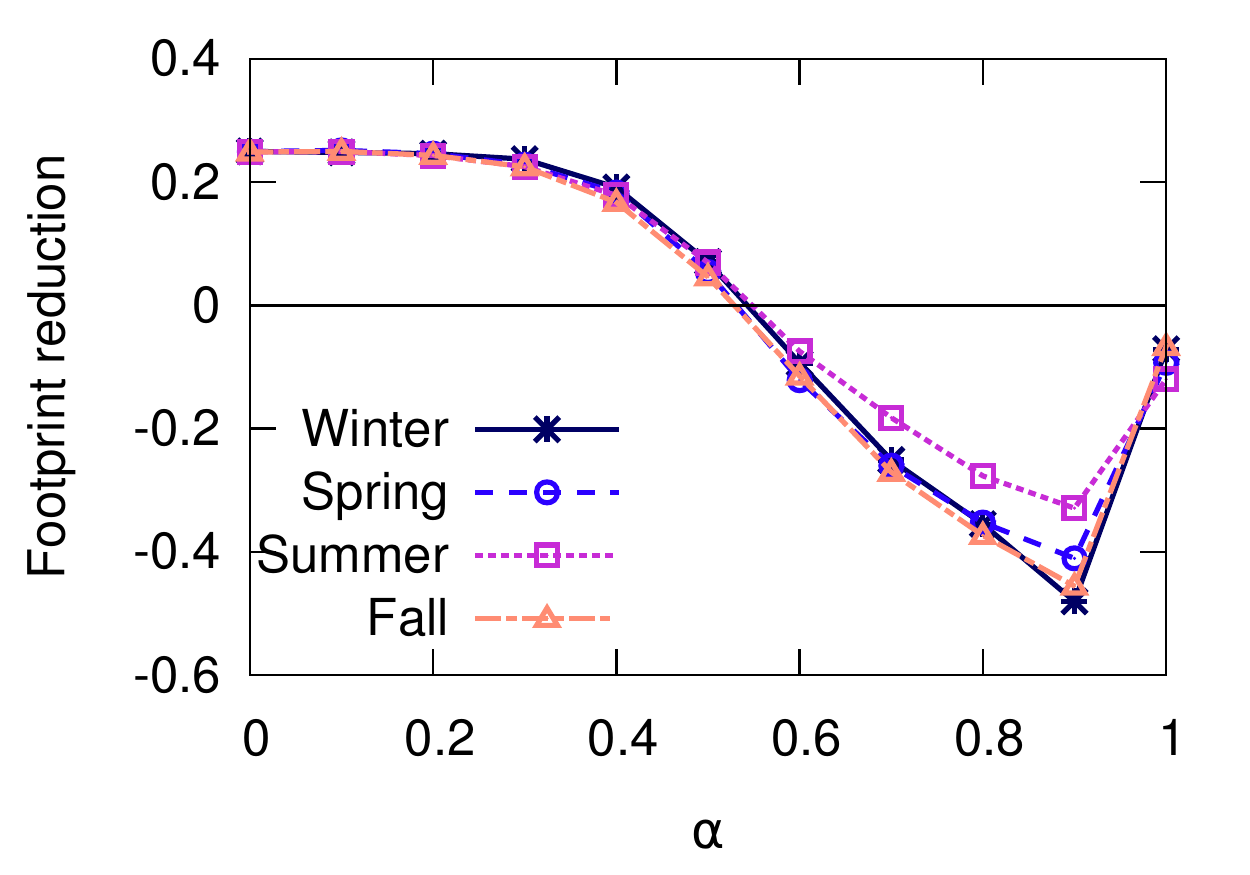}}
 \subfloat[NbSC]{\label{fig:footprintStaticNbsc}\includegraphics[width=0.3\linewidth]{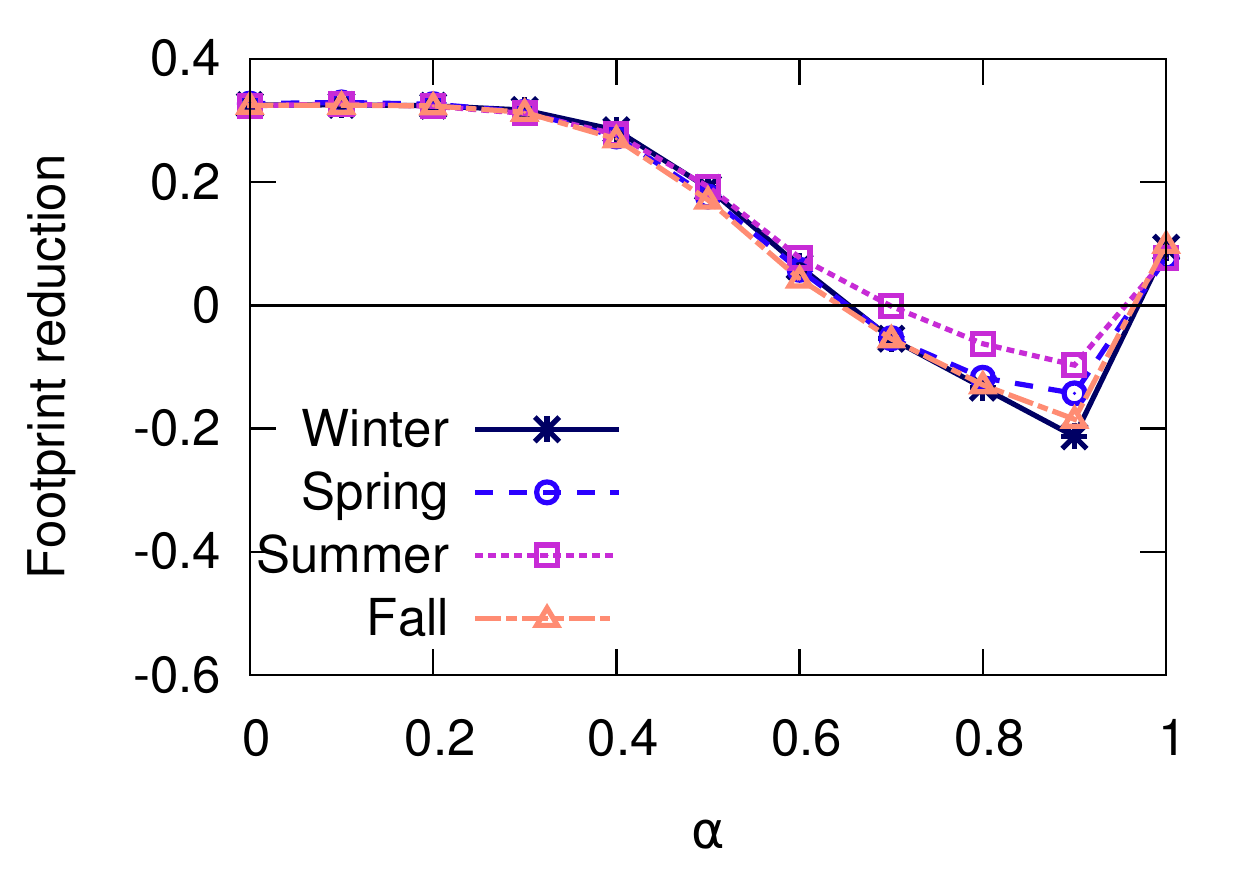}}
 \caption{Varying $\alpha$ of \rePGBR{} routing and evaluating its impact on footprint reduction for scenario A.}
 \label{fig:footprintStatic}
\end{figure*}

\begin{figure*}[!t]
 \centering
 \subfloat[ALL]{\label{fig:footprintRandomAll}\includegraphics[width=0.3\linewidth]{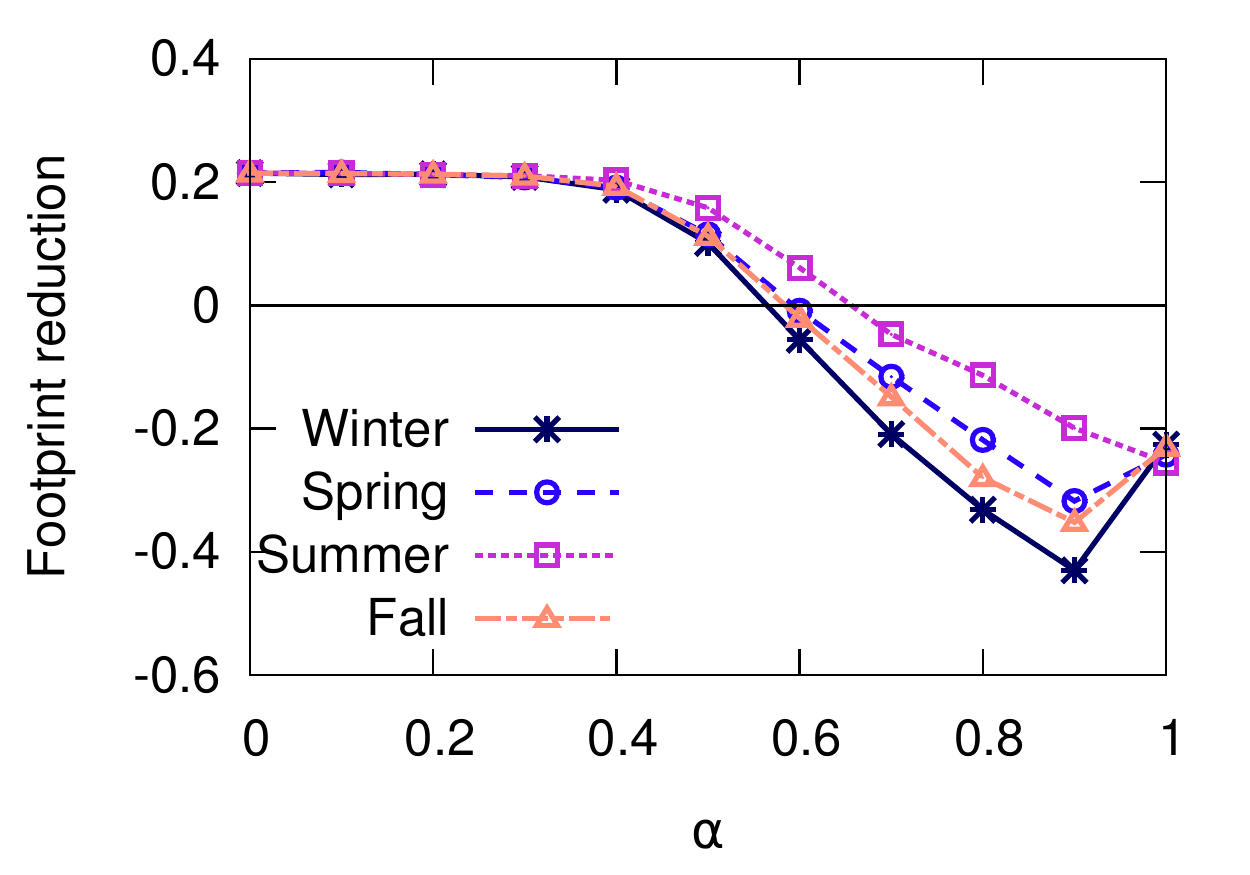}}
 \subfloat[Cachedbit]{\label{fig:footprintRandomCachedbit}\includegraphics[width=0.3\linewidth]{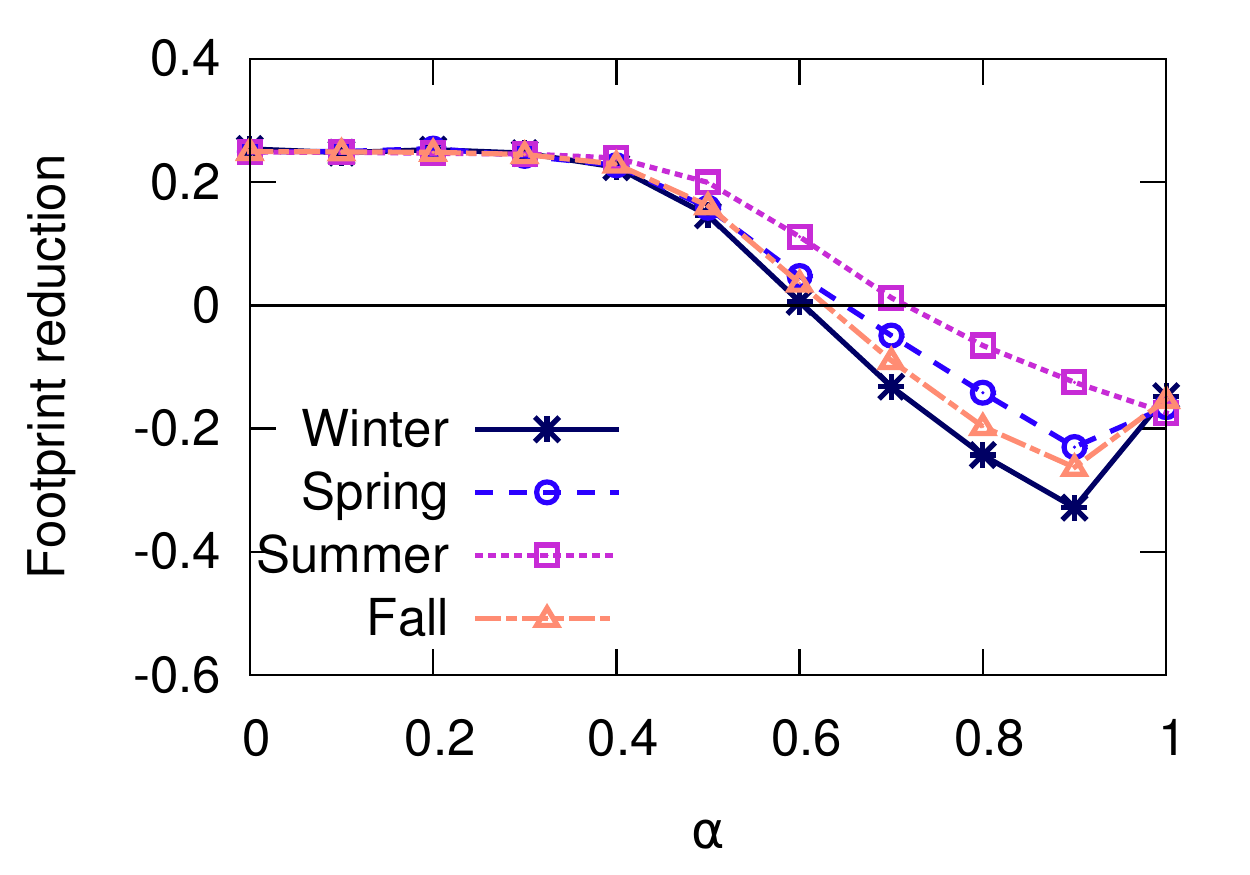}}
 \subfloat[NbSC]{\label{fig:footprintRandomNbsc}\includegraphics[width=0.3\linewidth]{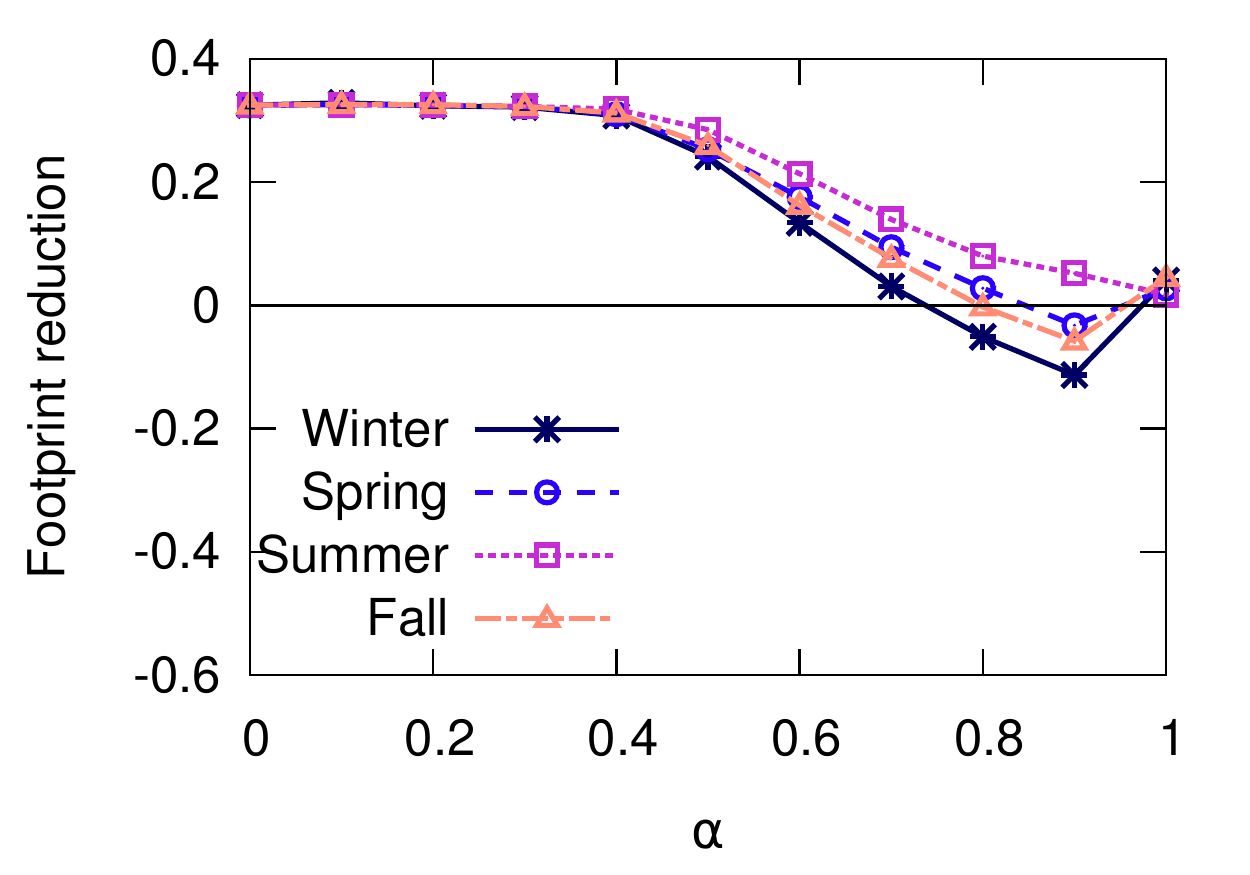}}
 \caption{Varying $\alpha$ of \rePGBR{} routing and evaluating its impact on footprint reduction for scenario B.}
 \label{fig:footprintRandom}
\end{figure*}

As described earlier, the benefit of the \rePGBR{} routing algorithm is in
the ability of manipulating $\alpha$ to suit any particular objective.
In our scenarios, we would ideally like to select an $\alpha$
that provides the greenest routes, without compromising on the caching
performance. Therefore, we have conducted tests to select the most
appropriated $\alpha$ for different seasons in the year. In total, we
investigated 4 weeks throughout the year, and the periods of these
weeks are as follows (Winter: 1st - 7th, January; Spring: 6th - 12th,
April; Summer: 11th - 27th, July; Fall: 21st - 27th, October). The
weeks of Spring and Fall have been selected at different times to
ensure that the weather patterns would not be similar. Fall is
on the verge of Winter while Spring is slightly milder.

We first evaluate the different caching strategies with
respect to varying $\alpha$ values for scenarios A and B. Below we will look more closely at
performance of \rePGBR{} in these conditions.

\begin{figure*}[!t]
 \centering
 \subfloat[ALL - Green/brown ratio]{\label{fig:greenRatioStaticAll}\includegraphics[width=0.3\linewidth]{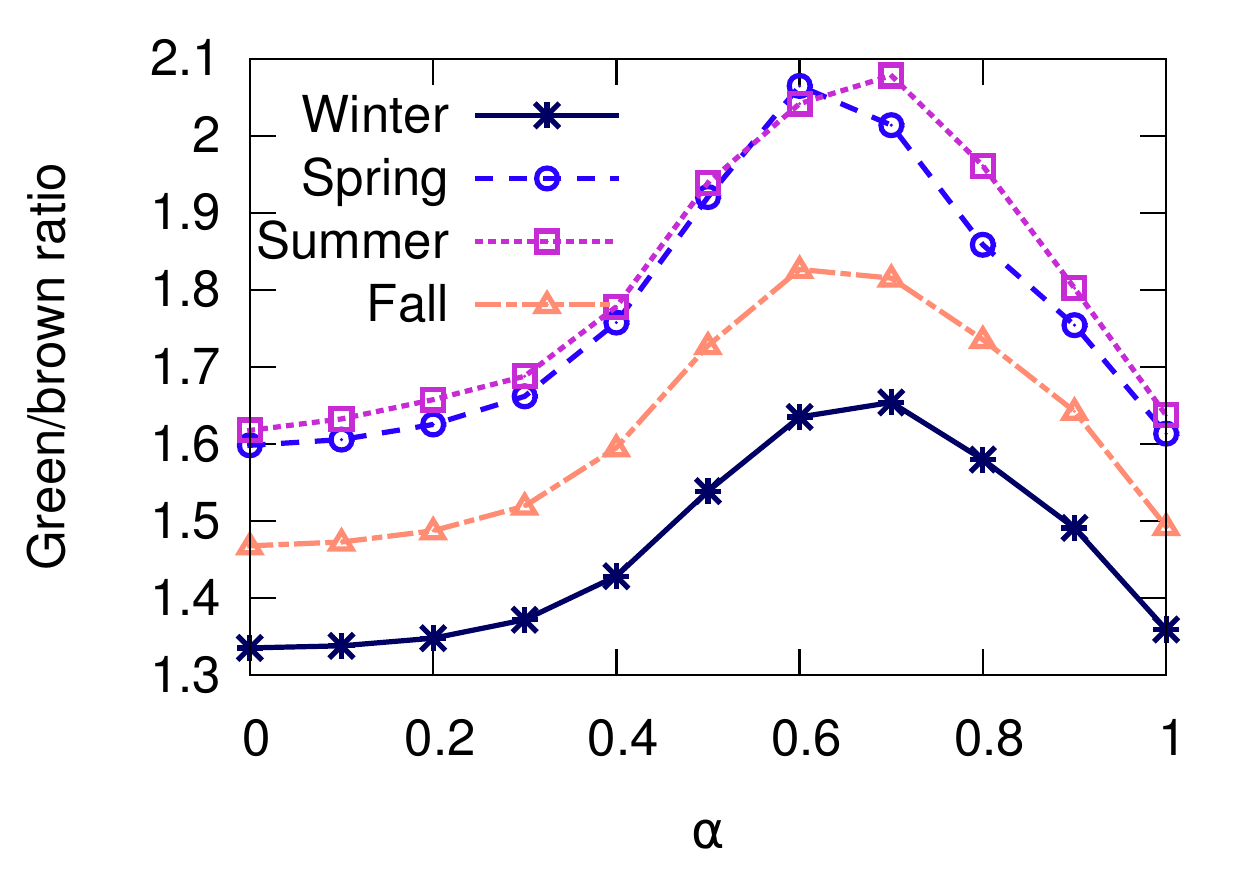}}
 \subfloat[Cachedbit - Green/brown ratio]{\label{fig:greenRatioStaticCachedbit}\includegraphics[width=0.3\linewidth]{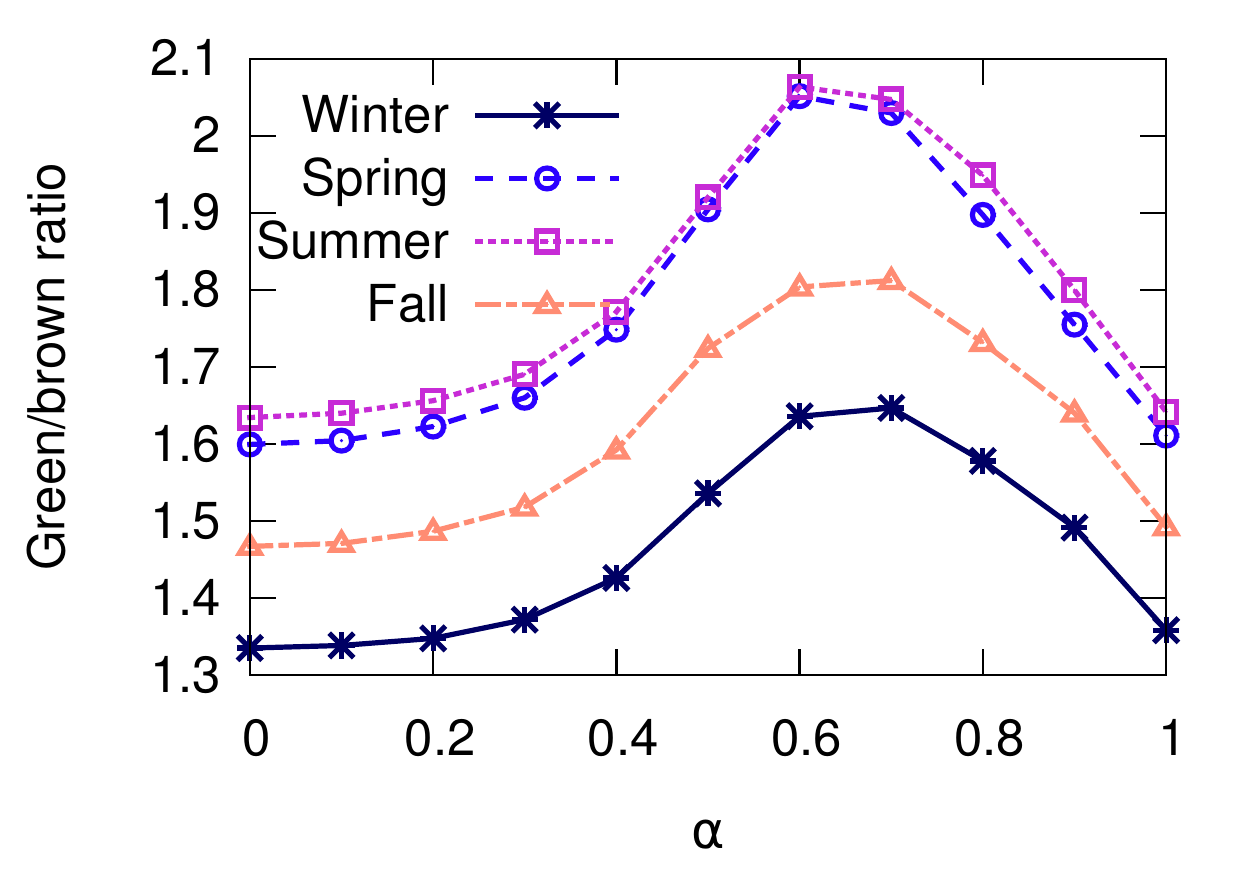}}
 \subfloat[NbSC - Green/brown ratio]{\label{fig:greenRatioStaticNbsc}\includegraphics[width=0.3\linewidth]{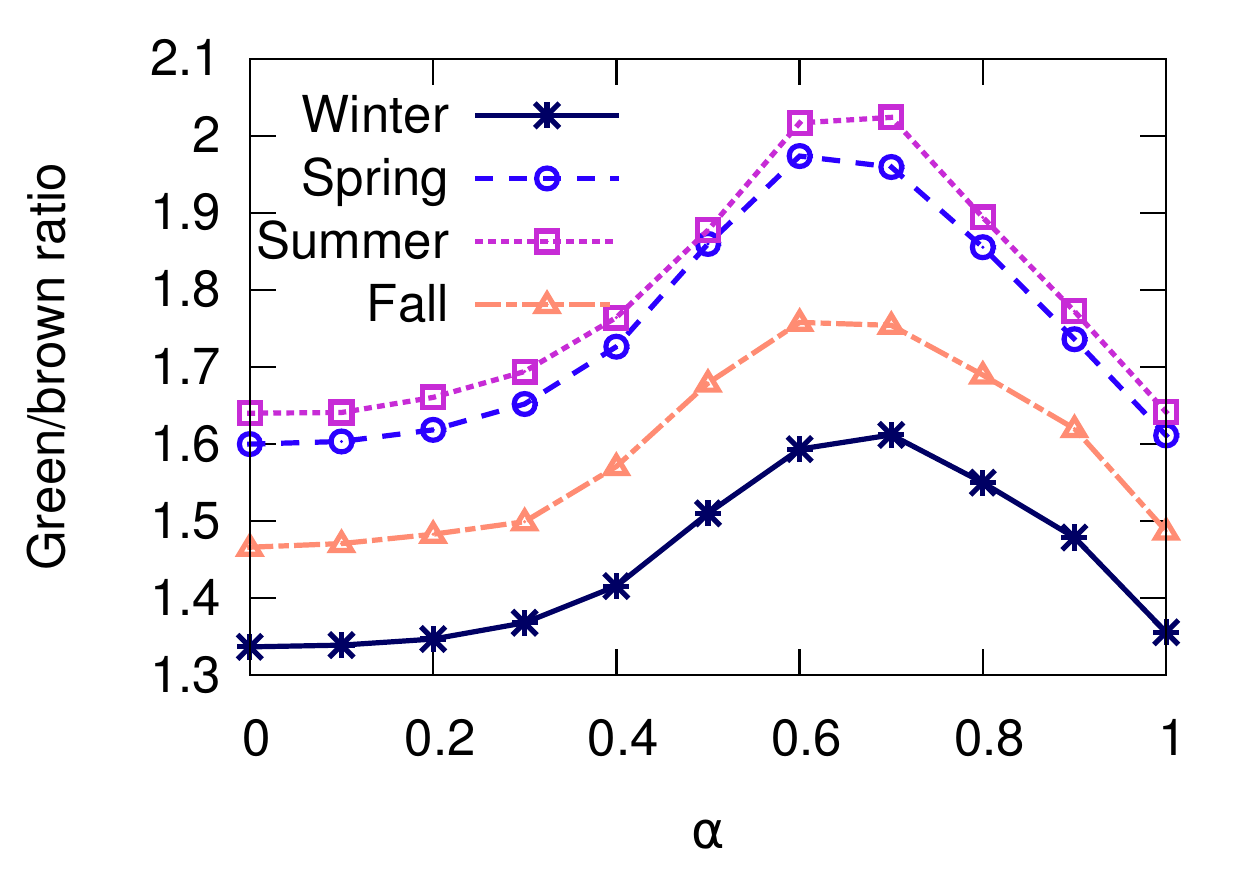}}\\
 \subfloat[ALL - Reduction of brown packets]{\label{fig:brownReductionStaticAll}\includegraphics[width=0.3\linewidth]{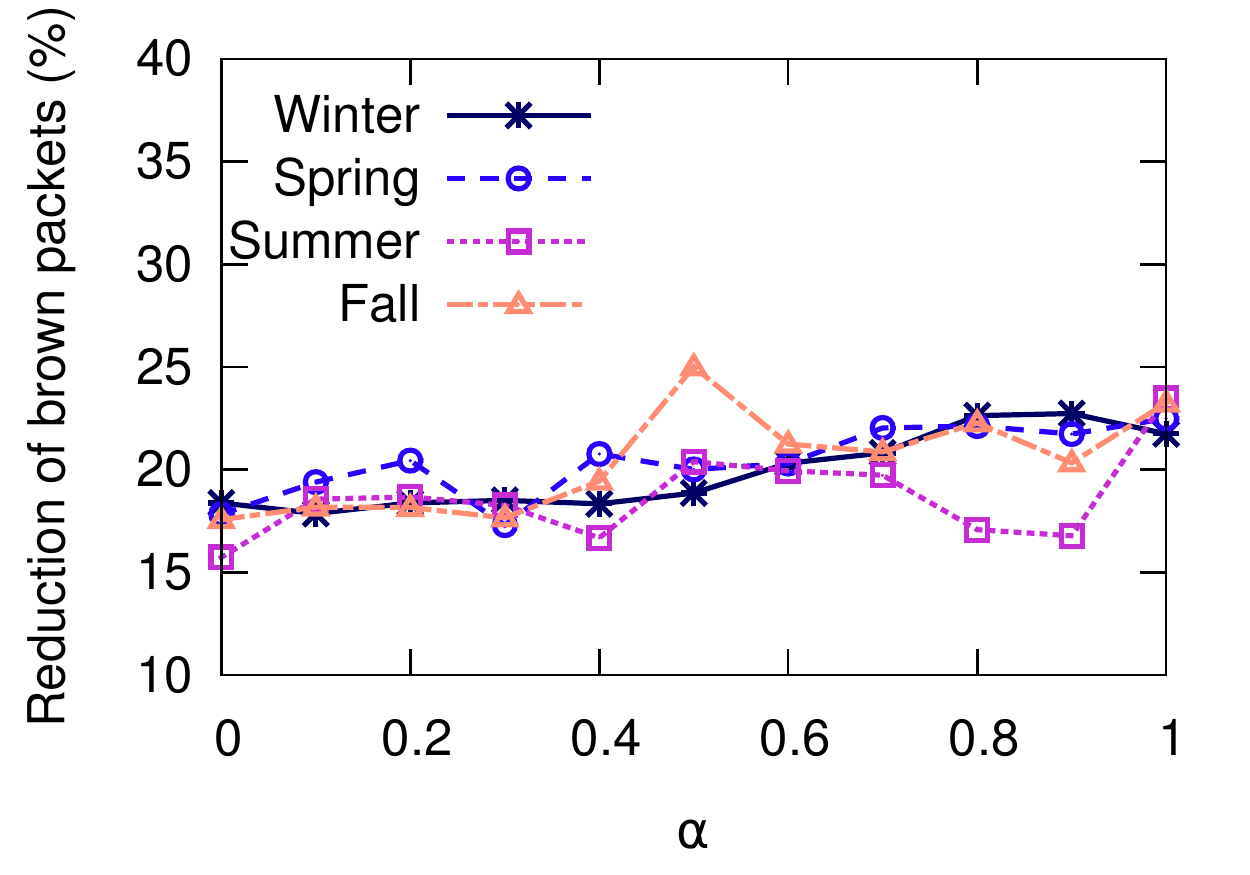}}
 \subfloat[Cachedbit - Reduction of brown packets]{\label{fig:brownReductionStaticCachedbit}\includegraphics[width=0.3\linewidth]{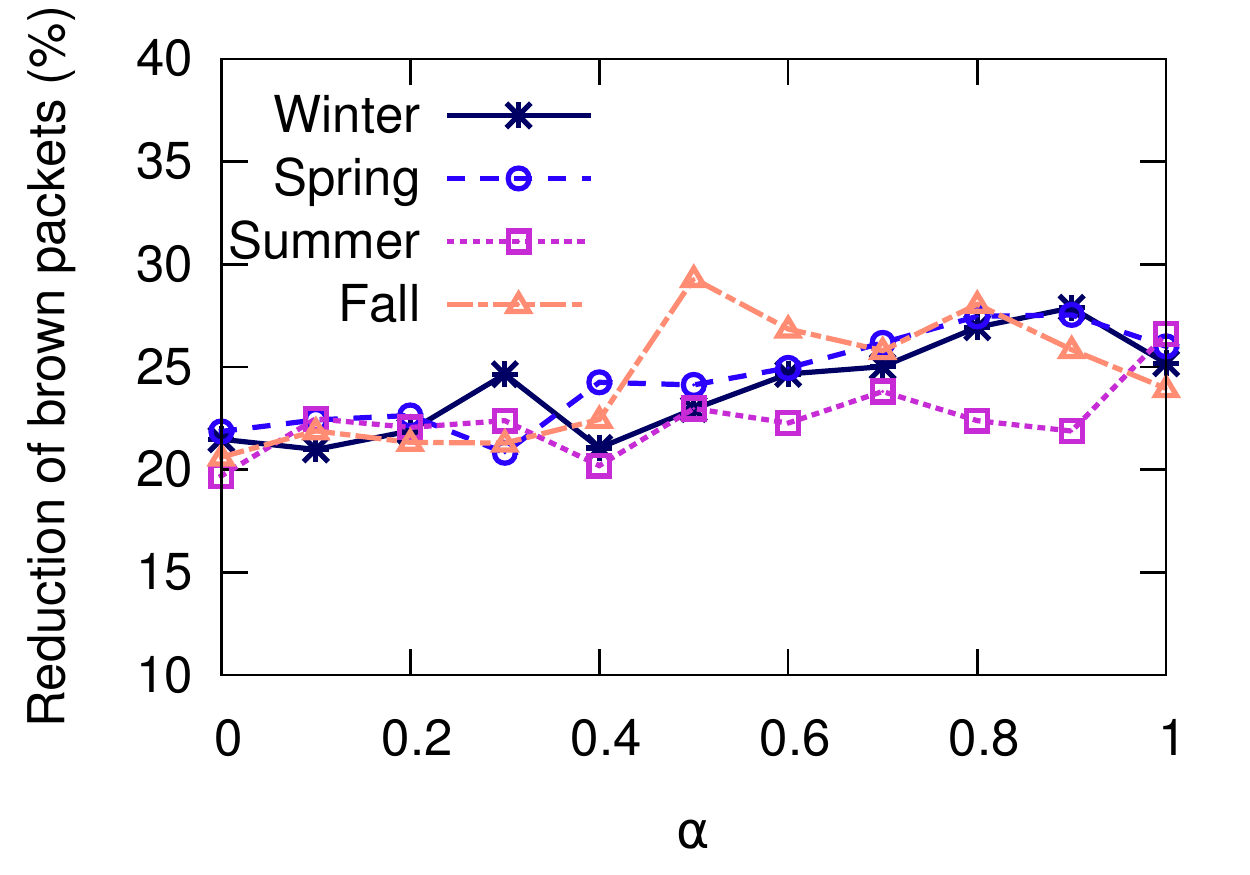}}
 \subfloat[NbSC - Reduction of brown packets]{\label{fig:brownReductionStaticNbsc}\includegraphics[width=0.3\linewidth]{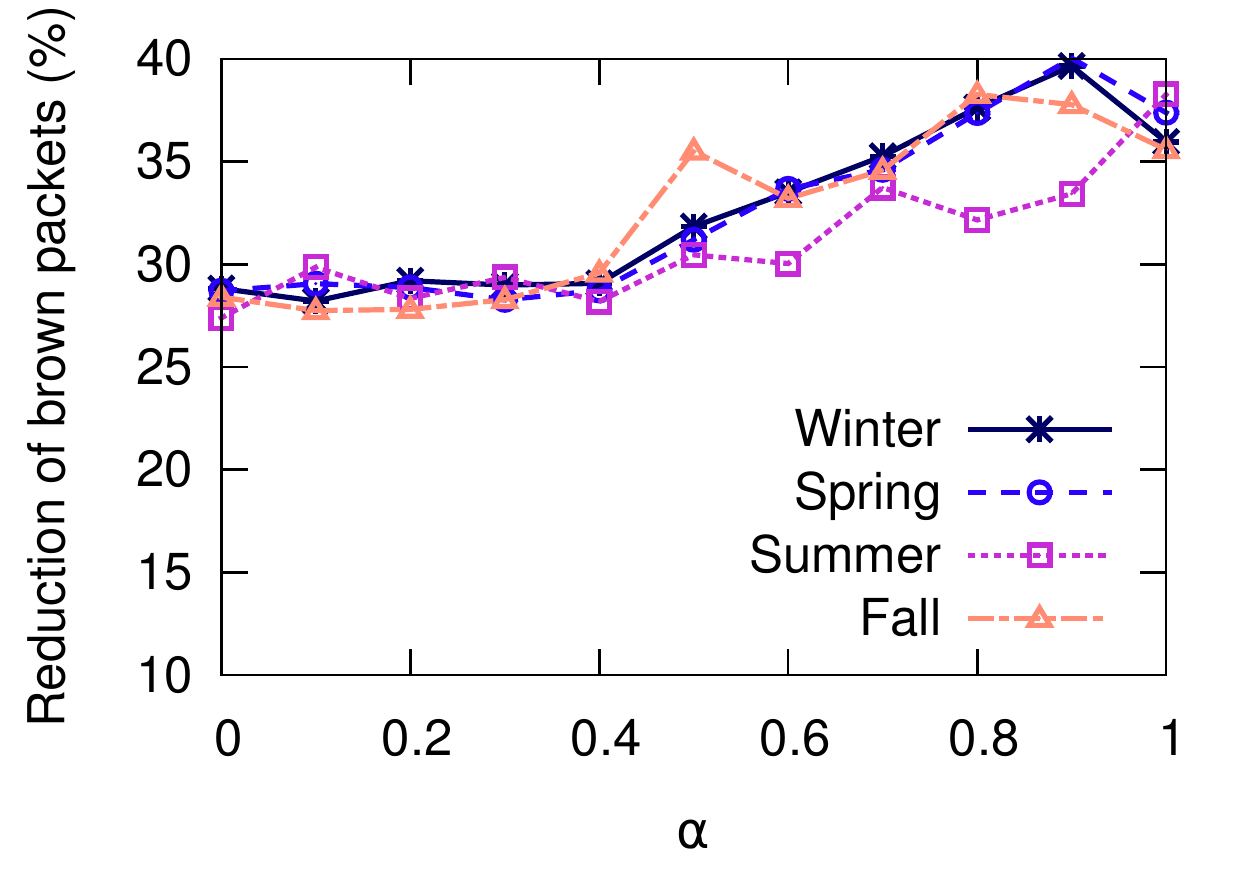}}
 \caption{Varying $\alpha$ of \rePGBR{} routing and evaluating its impact on green/brown ratio and reduction of brown packets for scenario A.}
 \label{fig:greenStatic}
\end{figure*}

We begin first by evaluating the different caching strategies with respect to varying $\alpha$ values. As previously mentioned, we focused on 5 metrics including
footprint reduction (e.g. \figurename{~\ref{fig:footprintStatic} and \ref{fig:footprintRandom}}), hit rate (e.g. \figurename{~\ref{fig:hitrateRandom}}),
green/brown ratio and the reduction of brown processed packets (e.g. \figurename{~\ref{fig:greenStatic}}), as well as the brown energy
savings of the network (e.g. \figurename{~\ref{fig:brownSavings_0_static} and \ref{fig:brownSavings_0_random}}). All the results are shown using \kW{5} wind turbines,
but with the nodes having different renewable energy architectures (scenarios A or B). Below we will look more closely at
the performance of the solution in these conditions.

\subsubsection{Footprint reduction}

We previously mentioned that the footprint reduction is a relevant metric for the ISP, showing the improvements in its internal traffic. \figurename{~\ref{fig:footprintStatic}} shows the
footprint reduction as a function of $\alpha$ for the four seasons and for each caching strategies. For all the caching strategies, we see that as $\alpha$ increases, the footprint reduction decreases. 
First the degradation is limited (when $\alpha \leq 0.3$). The reason for this positive correlation is because higher $\alpha$ value leads to longer paths along routers with high
green ratio, which further increases the footprint of the network. 
The results also show NbSC is far more superior than Cachedbit and ALL. The same behavior is observed for scenario B and \figurename{~\ref{fig:footprintRandom}} confirms this. 

In particular, after $\alpha$ reaches 0.4, the footprint reduction drops significantly. When the footprint reduction reaches zero, this means the paths are so
long that caching strategies can no longer reduce intra-network traffic. Obviously, caching still happens, but the net benefit is
zero. From the figure, we can see this threshold is about $0.8$ for NbSC (e.g. \figurename{~\ref{fig:footprintRandomNbsc}})
and lower for the others (e.g. \figurename{~\ref{fig:footprintRandomAll} and \ref{fig:footprintRandomCachedbit}}). 
The degradation of the footprint reduction varies with the seasons but the behavior is similar. 
Although the behaviour is similar, we can see that for all caching strategies there are improvements in footprint reduction in the warm seasons.
This improvement is visible when $\alpha$ is greater than 0.6 and 0.4 for scenario A and B, respectively.
An increase of the footprint reduction happens when $\alpha=1$, due to inefficient route discoveries that are not able to update routing table information 
with the latest renewable energy values. As a consequence, requests are served using  outdated paths increasing slightly the chances of a cache hit.

As we mentioned before, we considered a variant of NbSC which selects
the neighbor router with the highest green ratio, instead of a random neighbor when multiple
matches are found. However, we found out that this modification has no
visible effect on the performance. This is mainly because routers are
co-located at POPs and thus share the same weather pattern among many
(but not all) neighbors.

We observe a large performance difference between NbSC and the others even when using only the shortest path (with $\alpha=0$; ALL and Cachedbit has a footprint reduction of 21\% and 25\%, respectively, while NbSC has a footprint reduction of 32\% for the static and random scenarios). The advantage of NbSC is at least 7\% higher than Cachedbit and increases with $\alpha$, up to 23\% and 18\%, respectively for scenarios A and B.

\subsubsection{Improving network's greenness}

\begin{figure}[!t]
 \centering
 \subfloat[No caching / ALL / Cachedbit]{\label{fig:nothing_static_5kw_brownSavings_0}\includegraphics[width=0.5\figurewidth]{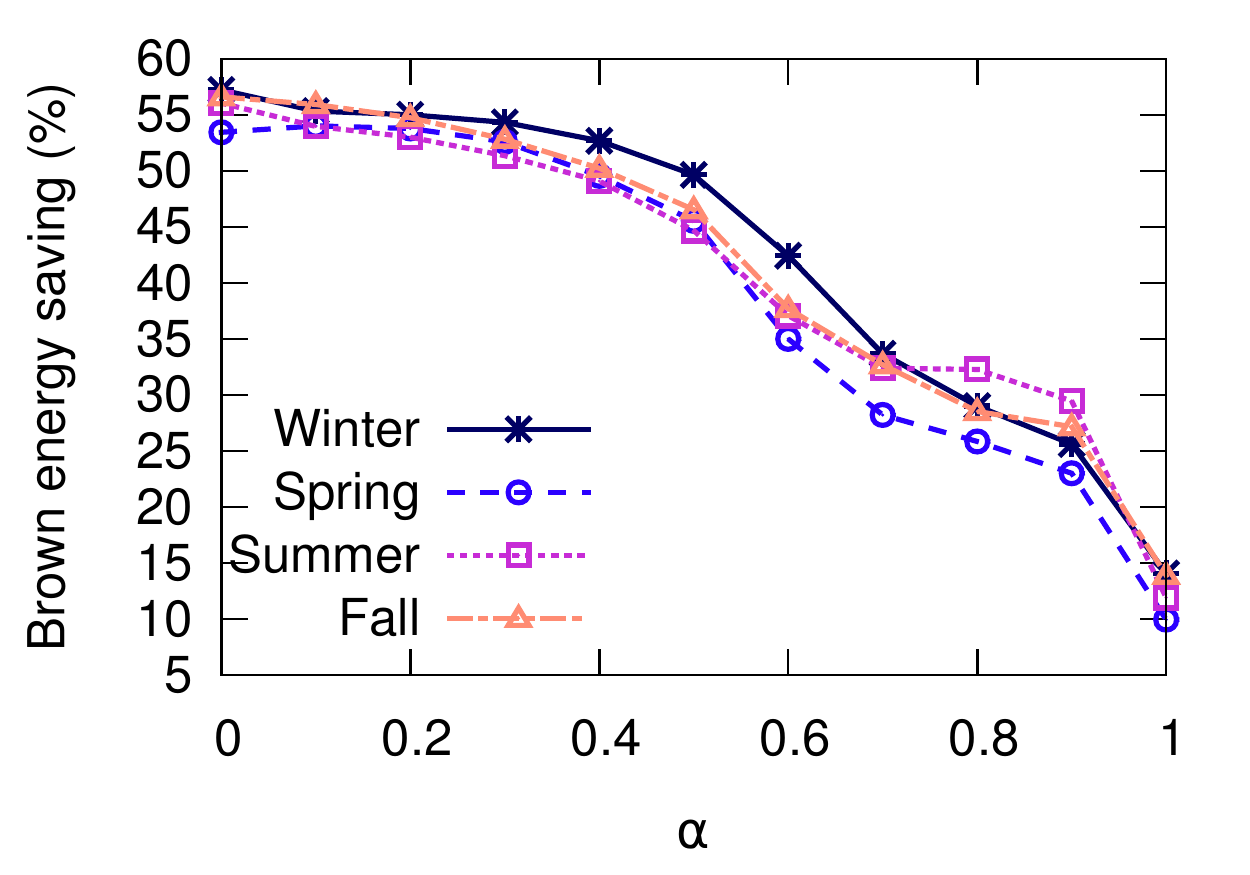}}
 \subfloat[NbSC]{\label{fig:nbsc_static_5kw_brownSavings_0}\includegraphics[width=0.5\figurewidth]{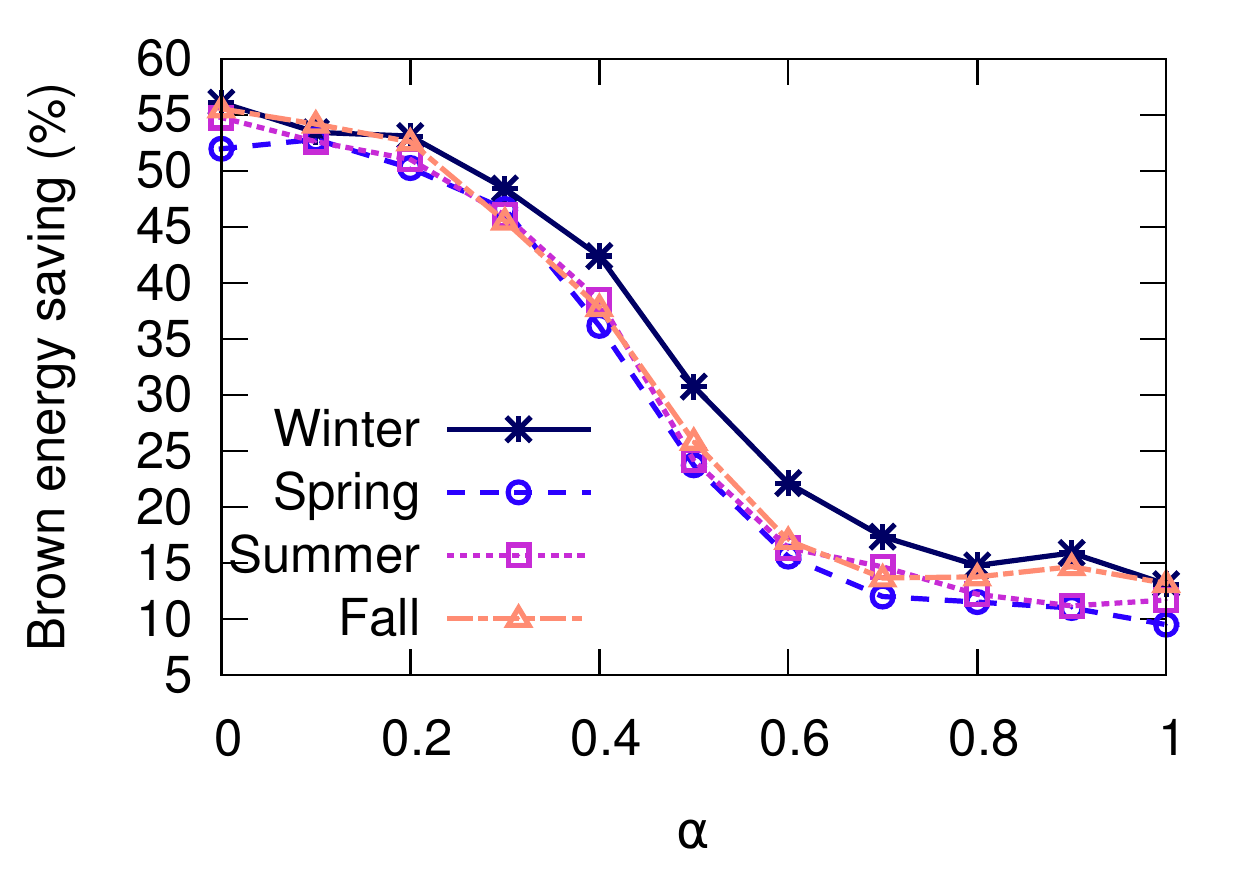}}
 \caption{Varying $\alpha$ of \rePGBR{} routing and evaluating its impact on brown energy savings for scenario A.}
  \label{fig:brownSavings_0_static}
\end{figure}

\begin{figure}[!t]
 \centering
 \subfloat[No caching / ALL / Cachedbit]{\label{fig:nothing_random_5kw_brownSavings_0}\includegraphics[width=0.5\figurewidth]{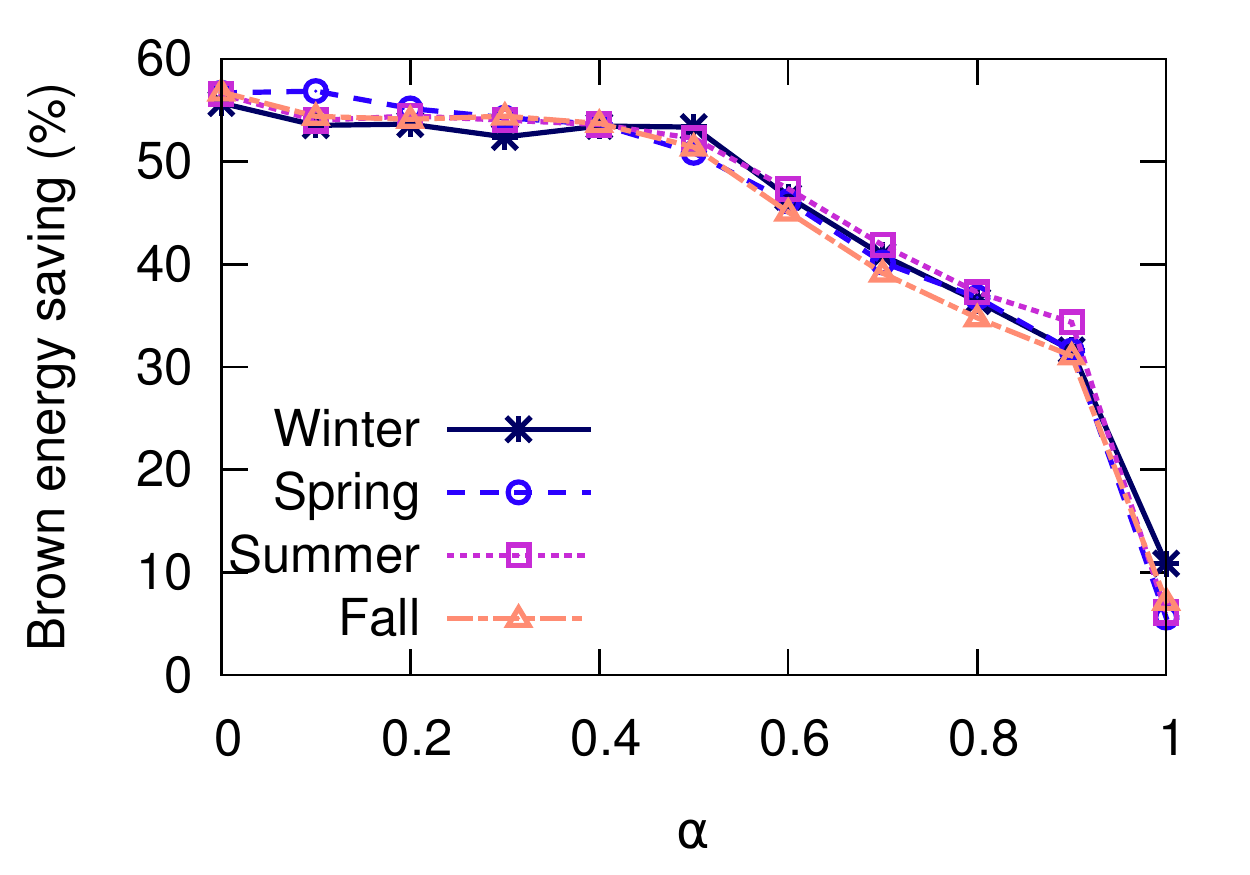}}
 \subfloat[NbSC]{\label{fig:nbsc_random_5kw_brownSavings_0}\includegraphics[width=0.5\figurewidth]{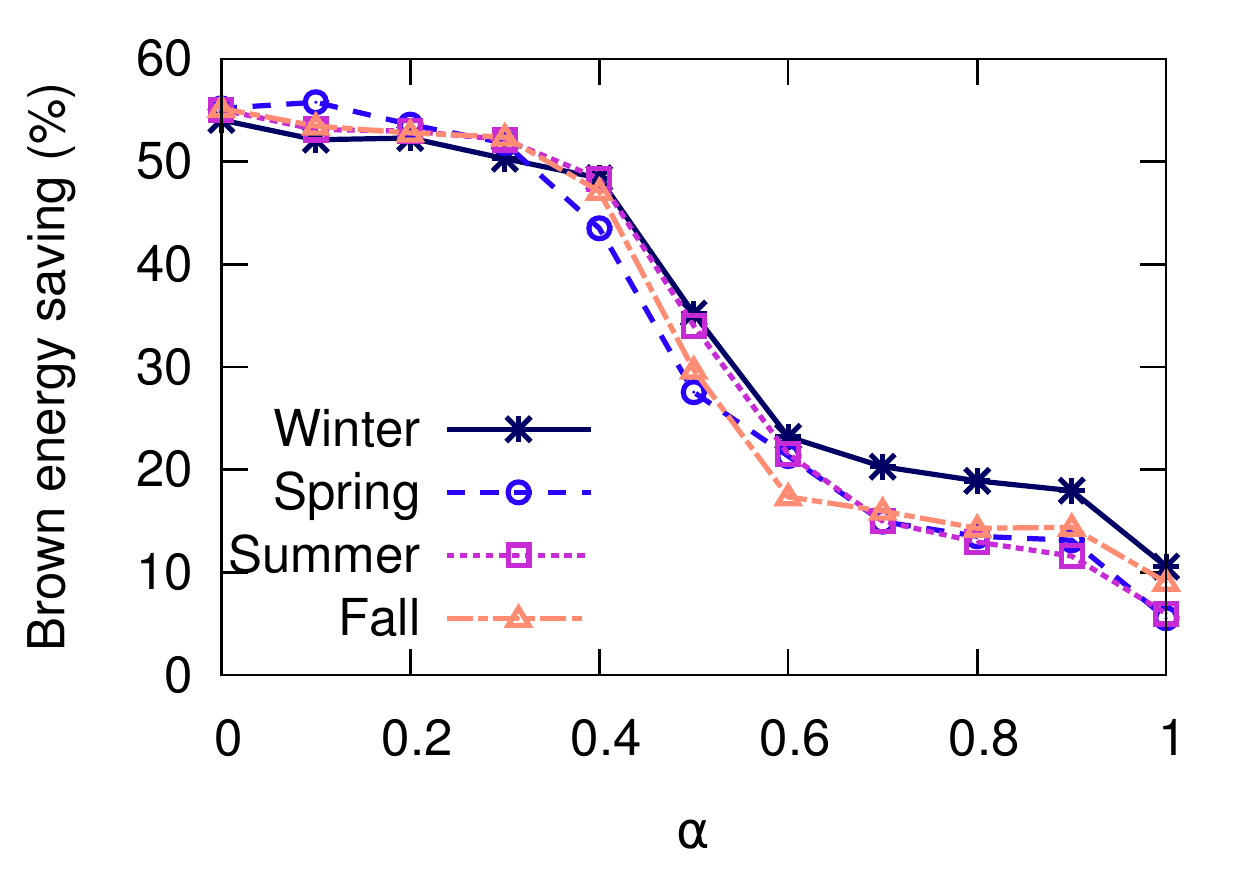}}
 \caption{Varying $\alpha$ of \rePGBR{} routing and evaluating its impact on brown energy savings for scenario B.}
 \label{fig:brownSavings_0_random}
\end{figure}

\figurename{~\ref{fig:greenStatic}} shows the ratio of green/brown packets and the reduction of
brown packets as a function of $\alpha$ for scenario A. In particular, \figurename{~\ref{fig:greenRatioStaticAll} - \ref{fig:greenRatioStaticNbsc}} show a
difference between the seasons for the green/brown ratio but very little between the caching strategies. This is due to the importance of \rePGBR{} route discovery to
enhance the green processing of the packets. In addition, we can see that the ratio is greater for the warm seasons (i.e. Summer and Spring) than for the cold seasons. One can see
that the performance of the warm seasons are quite similar compared to those for the cold seasons. As explained in the appendix, the warm seasons tend to be more
reliable (the impact of the solar energy is higher than the wind energy) and predictable (diurnal phases) which leads the routing protocol to maximize the discovery of highly green
routes.
\figurename{~\ref{fig:greenRatioStaticAll} - \ref{fig:greenRatioStaticNbsc}} show that the ratio of green/brown packets slowly increases with $\alpha$. 
As shown in the figures, the significant improvement (peak) starts at approximately 0.4 and ends at 0.9. Therefore, this is the range of $\alpha$ values that maximize the green
processing of packets in the network. However, the ratio of green/brown packets is insufficient without showing the reduction of brown packets as a function of $\alpha$, which are shown
in \figurename{~\ref{fig:brownReductionStaticAll} - \ref{fig:brownReductionStaticNbsc}}. The results show that the number of brown packets almost linearly decreases as 
$\alpha$ increases, which comforts the fact that the network's greenness increases when $0.4 \leq \alpha \leq 0.9$. Finally, the results also show that NbSC outperforms
other caching strategies for every $\alpha$.

Experimental results show similar trends for scenario B, that differ only from the green/brown packet ratio induced by the scenario definition. 
In the appendix, we described how the capacities influence on the greenness of the routers, where certain routers are more green that others irrespective 
of the weather conditions. This in turn limits the variety of routes discovered by \rePGBR{}.

Using the ability of \rePGBR{} to maximize the discovery of green routes, could leave to a number of routers left unused. \figurename{~\ref{fig:brownSavings_0_static} and 
\ref{fig:brownSavings_0_random}} shows the brown energy savings that could be achieved if these routers were powered-off. The best savings are obtained when
\rePGBR{} goes for the shortest routes, and decreases as $\alpha$ increases. The performances of ALL, Cachedbit and no caching are similar because all the routers discovered by 
decreasing $\alpha$ are used independently of the caching strategy. However, NbSC presents a significant disadvantage where the brown energy savings are reducing much faster
with regards to $\alpha$. For Cachedbit, the brown energy savings are quite stable until $\alpha$ reaches $0.4$, while NbSC maintains high energy savings only until $0.2$. 
To summarize, the brown energy savings of the network varies between 10\% and 55\% which could lead to significant improvement of the network's greenness. For scenario B,
the values of $\alpha$ that maintain high energy savings are larger than for scenario A. For instance, Cachedbit is able to maintain brown energy savings over 50\% when
$\alpha$ is lower or equal to $0.5$ and NbSC to $0.3$ for all seasons. As shown in \figurename{~\ref{fig:brownSavings_0_static} and \ref{fig:brownSavings_0_random}}, the impact of
seasonal weather conditions is not significant on the brown energy savings, because these savings are highly dependent on \rePGBR{} route discoveries.

\subsubsection{Hourly performance}

\begin{figure*}[!t]
 \centering
 \subfloat[Footprint reduction]{\label{fig:intervals_footprint}\includegraphics[width=0.25\linewidth]{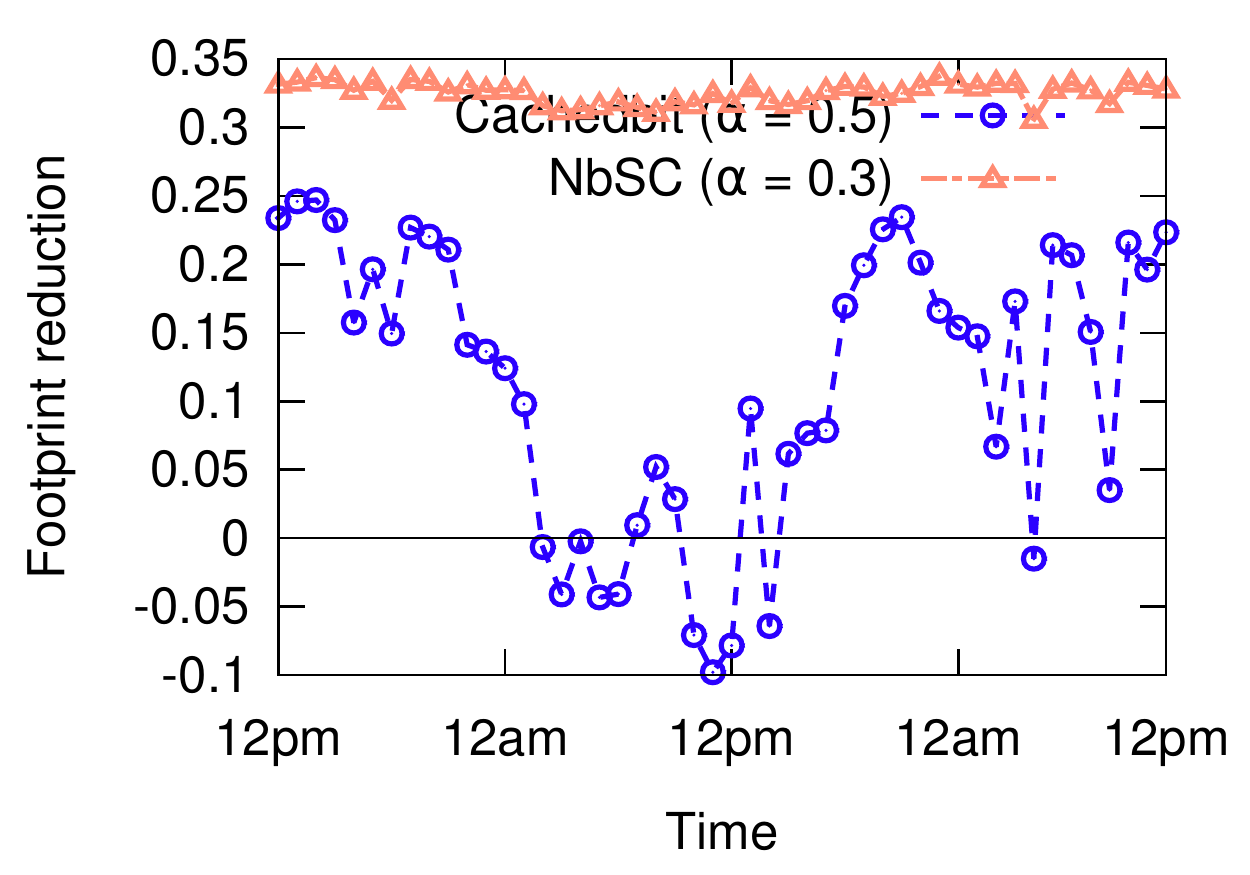}}
 \subfloat[Brown energy savings]{\label{fig:intervals_brownSavings}\includegraphics[width=0.25\linewidth]{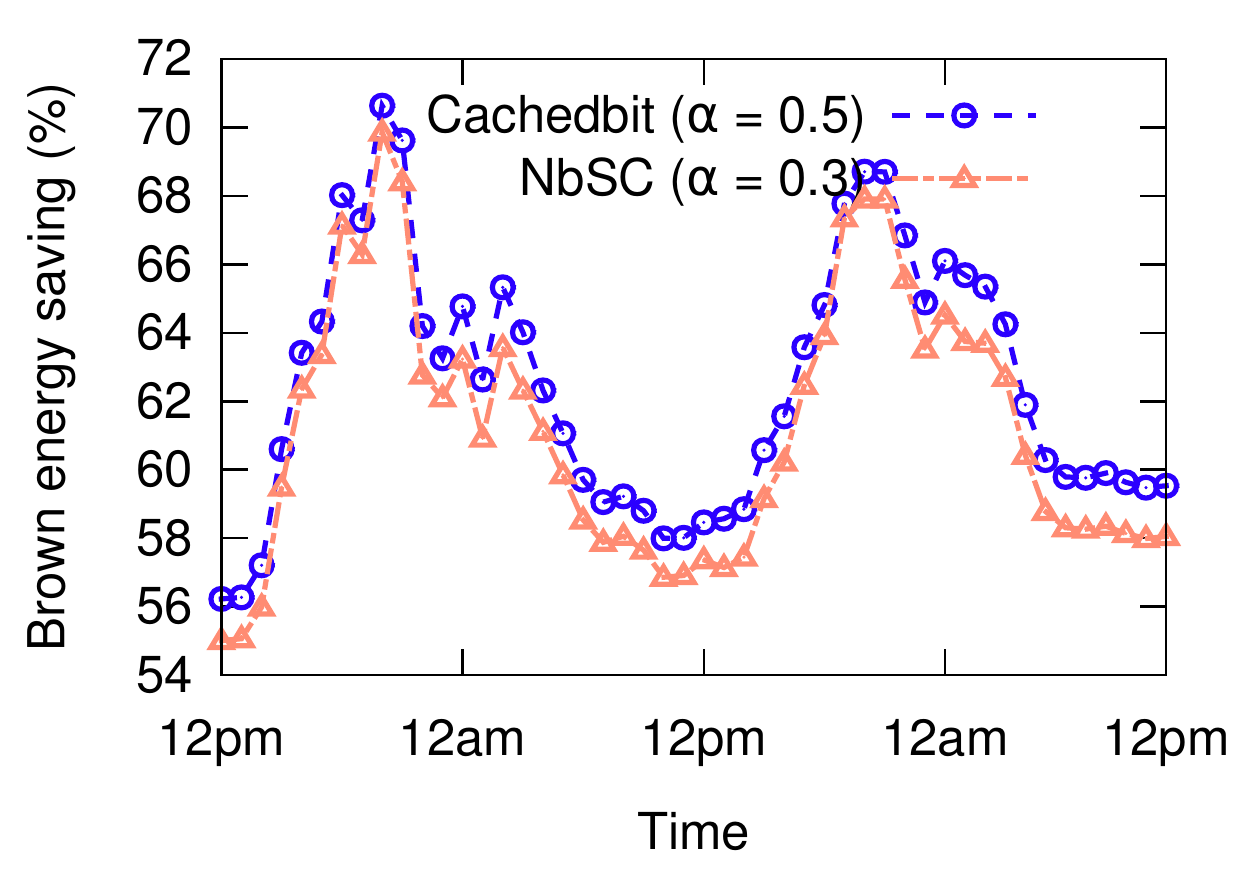}}
 \subfloat[Green/brown ratio]{\label{fig:intervals_greenRatio}\includegraphics[width=0.25\linewidth]{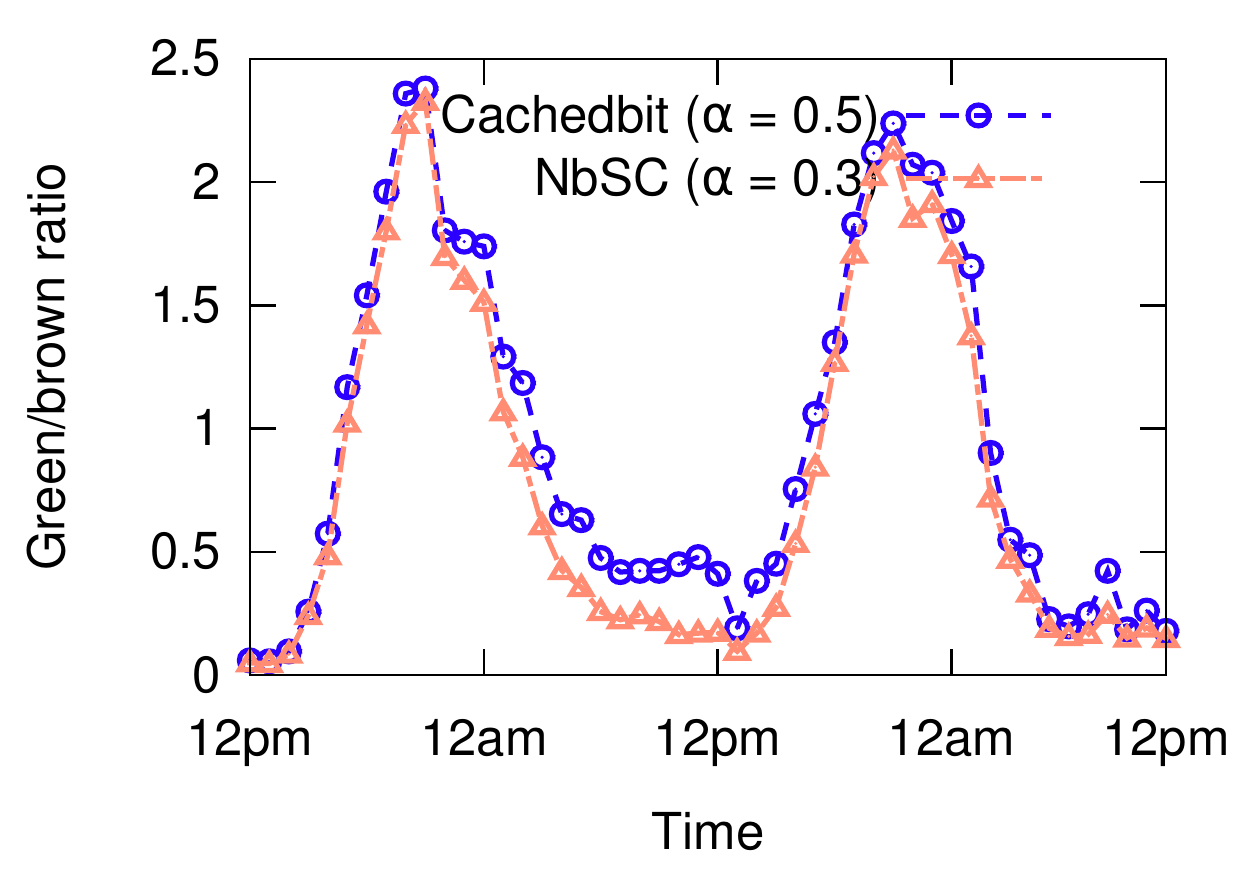}}
 \subfloat[Reduction of brown packets]{\label{fig:intervals_reductionPackets}\includegraphics[width=0.25\linewidth]{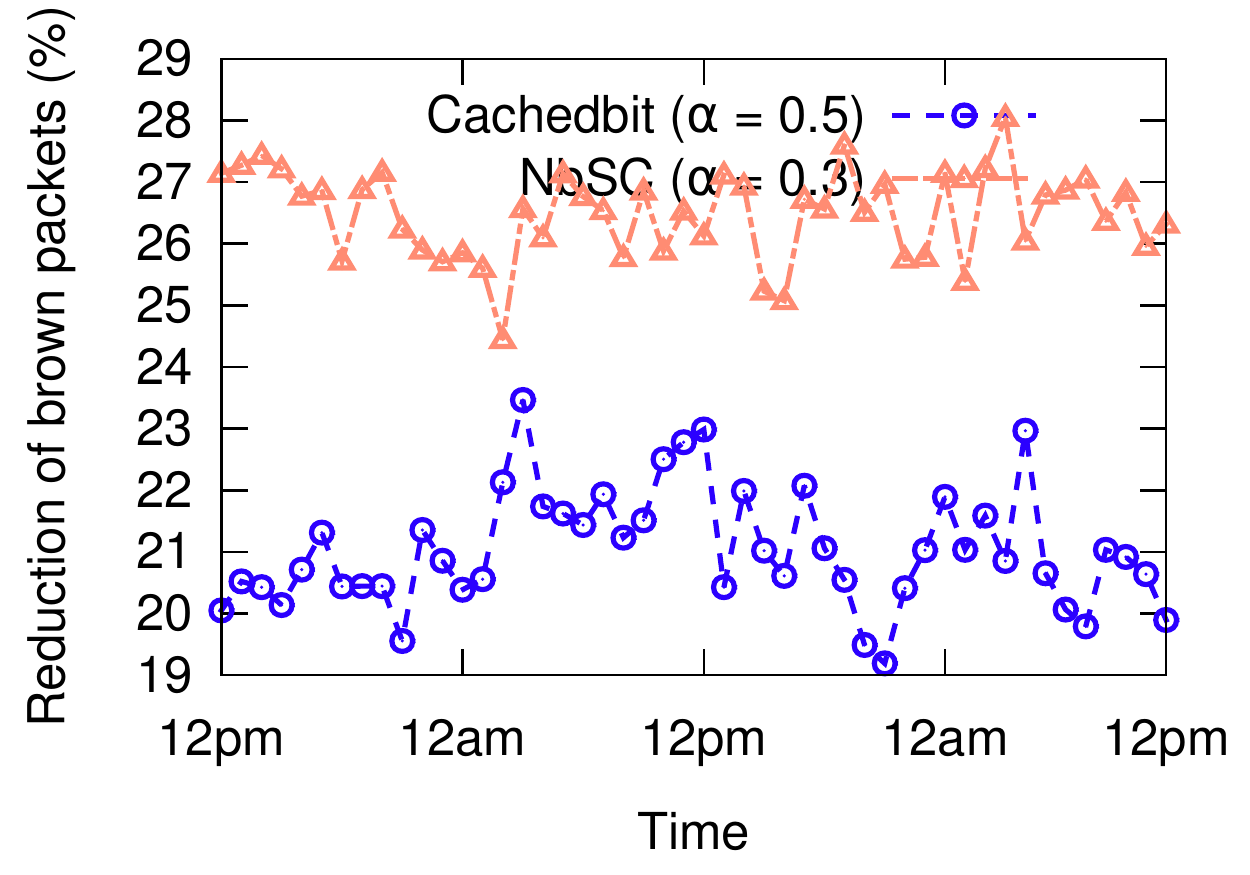}}
 \caption{Performances of Cachedbit and NbSC with respect to time during the two first days of Spring for scenario B.}
 \label{fig:intervals}
\end{figure*}

\figurename{~\ref{fig:intervals}} shows the 4 metrics described above for Cachedbit and NbSC as a function of time for the first 48 hours of Spring for scenario B. 
The $\alpha$ values used for each caching strategy correspond to the greenest $\alpha$ maintaining brown energy savings above 50\%, 
which are respectively $0.5$ and $0.3$ for Cachedbit and NbSC (\figurename{~\ref{fig:brownSavings_0_random}}).

In detail, \figurename{~\ref{fig:intervals_footprint}} shows the footprint reduction of Cachedbit and NbSC. The results of ALL behave 
similarly to Cachedbit's, although yield less gains, and as a consequence have been omitted. The results
are showing the high stability of NbSC which maintains a high footprint reduction of 32\% while Cachedbit's performance are highly variable throughout the days. The performances
of the two caching strategies are similar for the brown energy savings (e.g. \figurename{~\ref{fig:intervals_brownSavings}}) and the green/brown ratio 
(e.g. \figurename{~\ref{fig:intervals_greenRatio}}) and distinctively show a trend that maps to the sun pattern (see appendix). 
In other words, the performances of the caching strategies drastically increase during
the sunny hours. As previously mentioned, the sunny hours bring stability in the route discovery, thus improving the solution's performances. 
Lastly, \figurename{~\ref{fig:intervals_brownSavings}} shows the reduction of brown packets as a function of time, where the daytime pattern is not visible, and performances
of Cachebit and NbSC are fluctuating respectively around $20.5\%$ and $26.5\%$.

\subsection{Impact of the solution on data centers}

\begin{figure*}[!t]
 \centering
 \subfloat[ALL]{\label{fig:hitRateRandomAll}\includegraphics[width=0.3\linewidth]{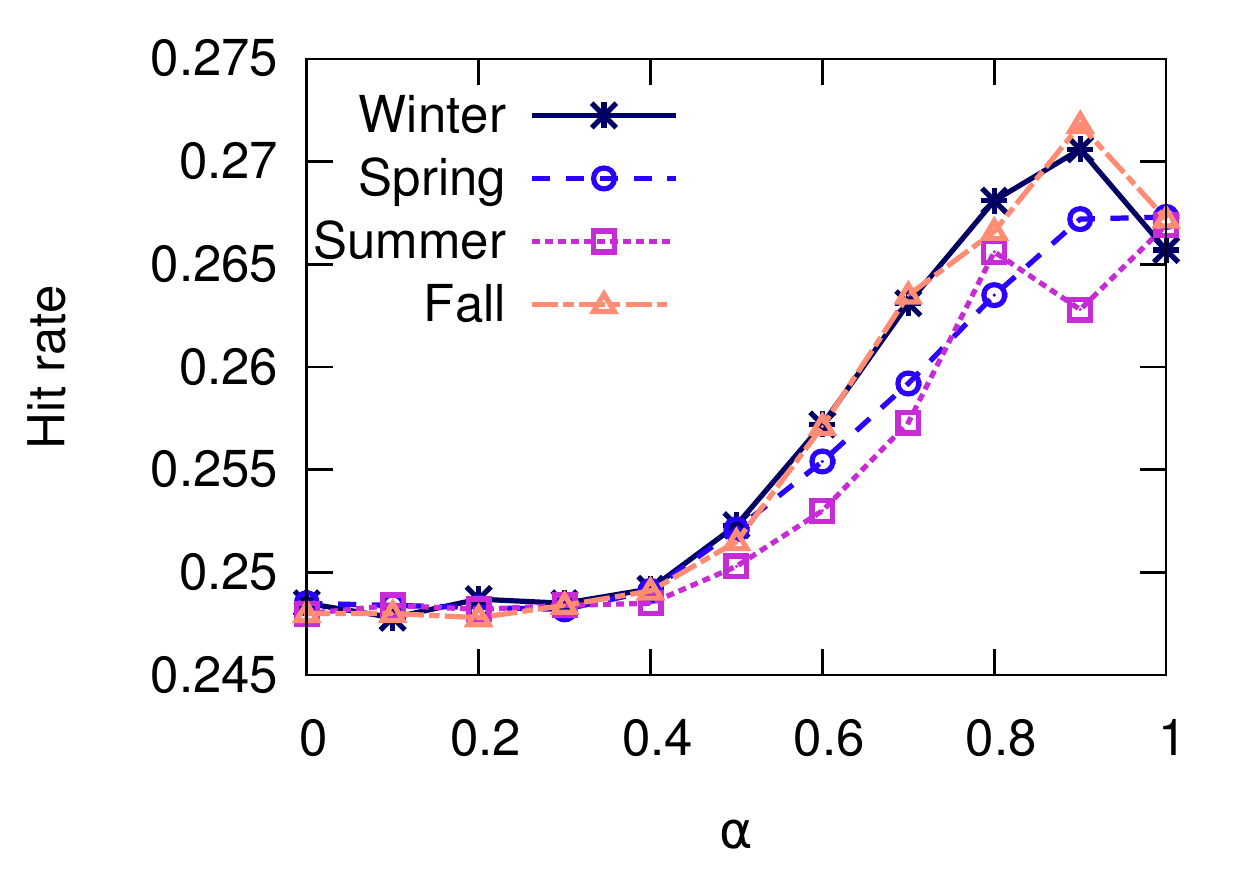}}
 \subfloat[Cachedbit]{\label{fig:hitRateRandomCachedbit}\includegraphics[width=0.3\linewidth]{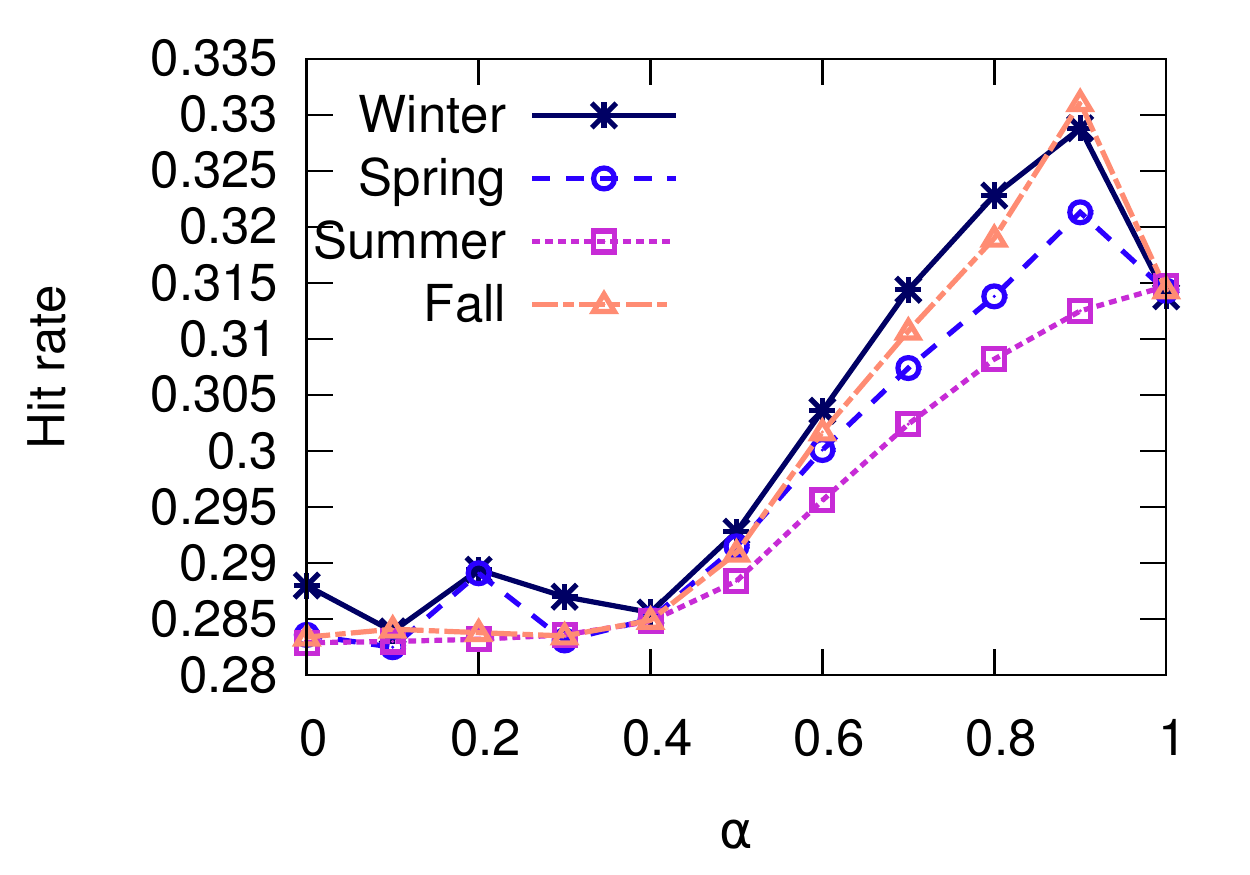}}
 \subfloat[NbSC]{\label{fig:hitRateRandomNbsc}\includegraphics[width=0.3\linewidth]{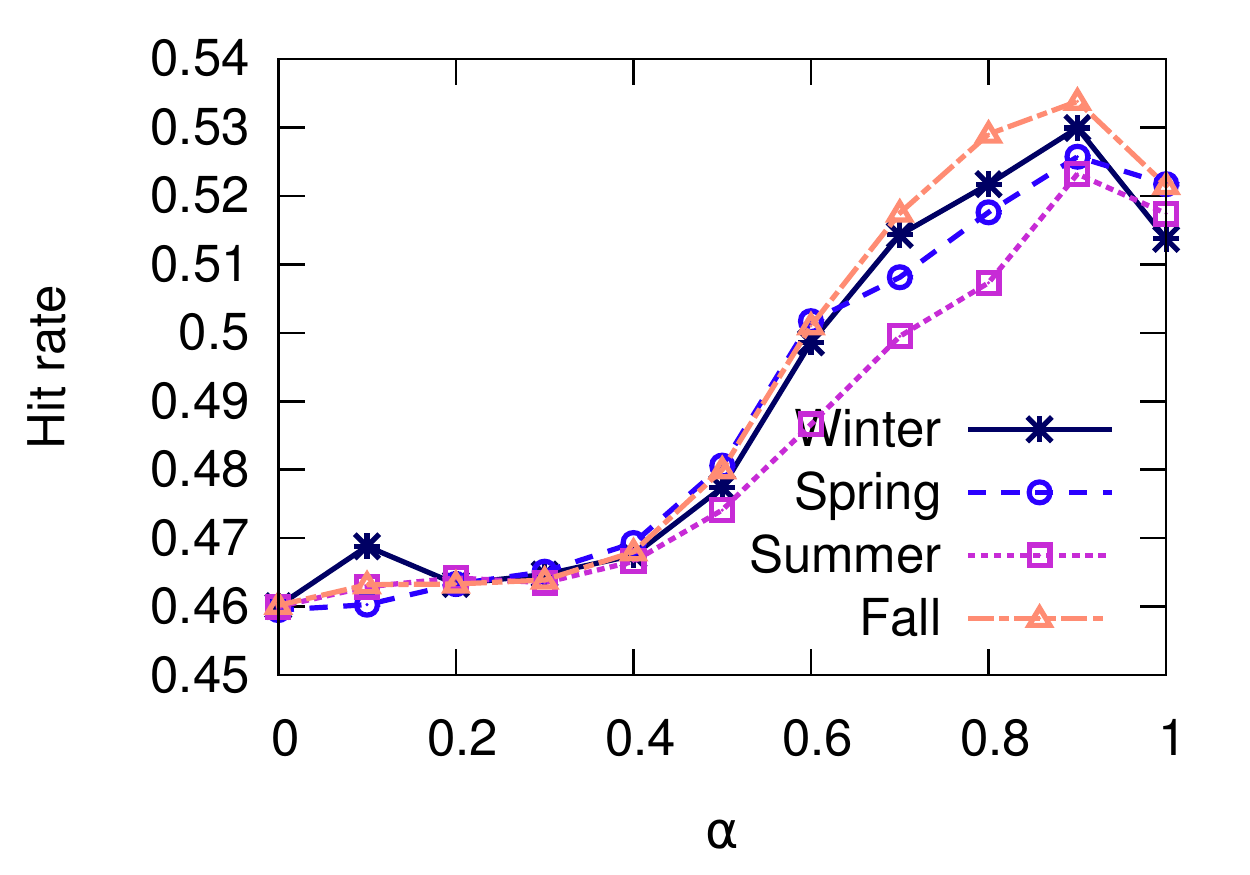}}
 \caption{Varying $\alpha$ of \rePGBR{} routing and evaluating its impact on hit rate for scenario B.}
 \label{fig:hitrateRandom}
\end{figure*}

\begin{figure}[!ht]
 \centering
 \includegraphics[width=\figurewidth]{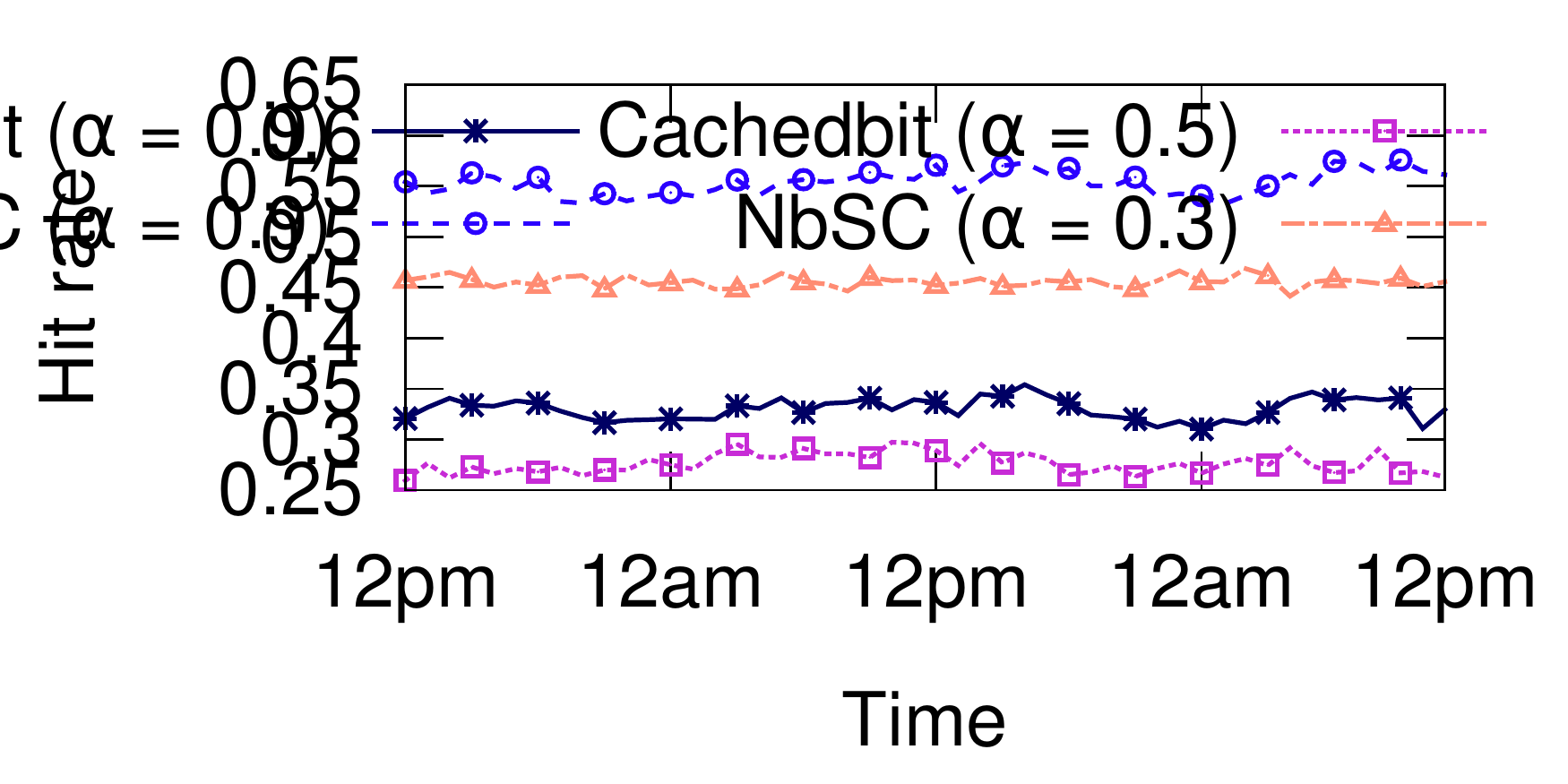}
 \caption{Hit rate performances of Cachedbit and NbSC with respect to time during the two first days of Spring for scenario B.}
 \label{fig:hitrate_interval_spring_random}
\end{figure}

In order to improve energy savings in the data centers, we assume that a non-negligible improvement could be obtained by reducing the number of requests directly served by the servers of
the data centers. Unlike the routers' energy consumption where the workload induced by traffic does not modify the energy consumed, the data centers energy consumption are known to
\emph{vary with their workloads}. 
Our solution, combining the energy savings at the ISP using greener routes and efficient caching strategies, could greatly improves the workload of the data centers by serving content 
cached in the ISP's routers, thus reducing their energy requirements (i.e. maximizing $\sigma_d(\alpha)$ defined in Section~\ref{sec:problemStatement}).

We see that as $\alpha$ increases, the hit rate also increases, as shown in \figurename{~\ref{fig:hitrateRandom}}. 
The reason for this positive correlation is because
higher $\alpha$ value leads to longer paths along routers with high green ratio, which further increases the possibility of cache-hit along the path. 
The results also show NbSC is far more 
superior than the Cachedbit and ALL, with a respective gains of over 15\% and 20\% improvement for scenario B. Experiments for scenario A have shown similar gains and, thus have been omitted.
Furthermore, the seasonal differences are also negligible, which shows that the effectiveness of the caching strategies defined by $\alpha$ would deliver predictable performances.

The hourly performances of Cachedbit and NbSC to improve hit rate are depicted in \figurename{~\ref{fig:hitrate_interval_spring_random}} for a selection of $\alpha$ values on 
the two first days of Spring. The $\alpha$ values have been selected to show the hit rate performances when (i) the greenness of the ISP network is at its peak (where
$\alpha$ equal to $0.3$ and $0.5$ for NbSC and Cachedbit, respectively) or when (ii) the hit rate performance is at the maximum (with $\alpha=0.9$ for both strategies). \figurename{~\ref{fig:hitrate_interval_spring_random}} shows that hit rate performances are very stable and do not vary according to the weather conditions. However, one can observe that 
Cachedbit is constantly performing worse than NbSC with a loss of 20\% on average in both cases (i and ii).

As a result, increasing renewable energy awareness in routing enhances caching performances of the network, 
thus resulting on a positive impact on the data centers' workload. This in turn could lead to data centers reducing their energy consumption for unused servers that hold contents which are already cached within the network. 

\subsection{Discussion} 

As stated in Section~\ref{sec:problemStatement}, the aim of the paper is to demonstrate that using renewable energy awareness in routing at the ISP's network could reduce the 
energy consumption from fossil fuels of the network and the data centers serving user's requests. The network operator, could with the choice of a single parameter $\alpha$ and a caching strategy, 
maximize the brown energy savings of the network from an end-to-end perspective (i.e. including data centers), without compromising the performance of end user's access to contents.

\begin{itemize}
 \item \textbf{User-observed content delivery performance:} In
     this paper, we assume that delivery performance is positively
     correlated with footprint reduction. 
 \figurename{~\ref{fig:footprintStatic} and \ref{fig:footprintRandom}}
 showed that the performance would decrease as the renewable energy
 awareness increases, and this is observable
 for all caching strategies. There are even $\alpha$ values where the
 performance is worse than not using any caching 
 (i.e. footprint reduction below 0). For all caching strategies, the
delivery performance is maximal when the energy awareness is nil. Fortunately, the performance is maintained for lower $\alpha$ values, 
 and using NbSC further maximizes the delivery performance.
 \item \textbf{Network greenness:} Maximizing network's greenness includes minimizing use of brown energy by powering-off unused routers and 
 reducing the number of brown packets. Unfortunately,
 these performances are not achievable using the same $\alpha$ values and caching strategies.
 On the one hand, in order to maximize the brown energy savings of the network, the smallest $\alpha$ values and Cachedbit are preferable. 
 On the other hand, maximal reduction of brown packets is obtained with high $\alpha$ values and NbSC. 
 Depending on the requirements of the network, the operator would have to carefully
 choose the appropriate combination to best fulfill the network's objectives.
 \item \textbf{Data center greenness:} Improving caching performances also improves the hit rate and thus helps reduce the workload on data centers. Increasing $\alpha$ improves the hit rate, while
 using NbSC always increases the performances by an additional 20\%.
\end{itemize}
 
Therefore, the choice of the caching strategy and 
$\alpha$ value could produce a positive trade-off, where benefits would be maximized.

%% file: conclusion.tex
The popularity of the Internet today has led to widespread deployment of ICT infrastructures, which is consuming considerable quantity of energy. In recent years, a new research 
initiative towards enhancing renewable energy aware ICT infrastructure has been proposed. In this paper, we propose a renewable energy aware CCN, where routers are powered directly 
by wind and/or solar energy. The proposed approach is composed of two components, which includes a novel gradient-based routing algorithm 
that discovers paths along routers powered by high 
renewable energy, followed by a caching strategy that stores contents along these discovered paths. This enables the routing protocol as well as caching strategy, to adapt to varying 
weather patterns that may affect energy that is used to power the routers. The results from an experimental testbed using real meteorological data, has shown that using the combined 
approach has resulted in increased consumption of renewable energy, where a combination between high renewable energy 
consumption and caching strategy can minimize the need of fossil fuels energies of the network and the data centers without compromising the users' experience.

%% file: renewableEnergy.tex
We use publicly available data from the National Renewable Energy Laboratory~\cite{nrel} to estimate the availability of wind and solar power in different parts of the U.S. 
Below we describe our weather data and present the methodology used to convert the meteorological wind and solar data, to power.

\subsection*{Wind energy profile}

\figurename{~\ref{fig:windSpeed}} shows a simplified version of the
annual wind speed in the U.S.
The darker zones have an
average wind speed between $4$ - $6~m/s$, the average wind speed for
the medium zones are between $6$ - $8~m/s$, and in the brighter zone,
the average speed is greater than $8~m/s$.

We consider two wind turbines produced by the Huaying
company~\cite{windTurbineGenerator}. The two models are ``HY30-AD11''
which is a \kW{30} wind turbine, and a smaller model ``HY5-AD5.6'' which
produces energy up to \kW{5.4} (the larger turbine has a similar size
to the turbine investigated in \cite{Liu_Lin_Wierman_Low_Andrew_2011}.)

\begin{figure}[!t]
  \centering
  \subfloat[Wind speed to power conversion]{\label{fig:windToPower}\includegraphics[width=.5\figurewidth]{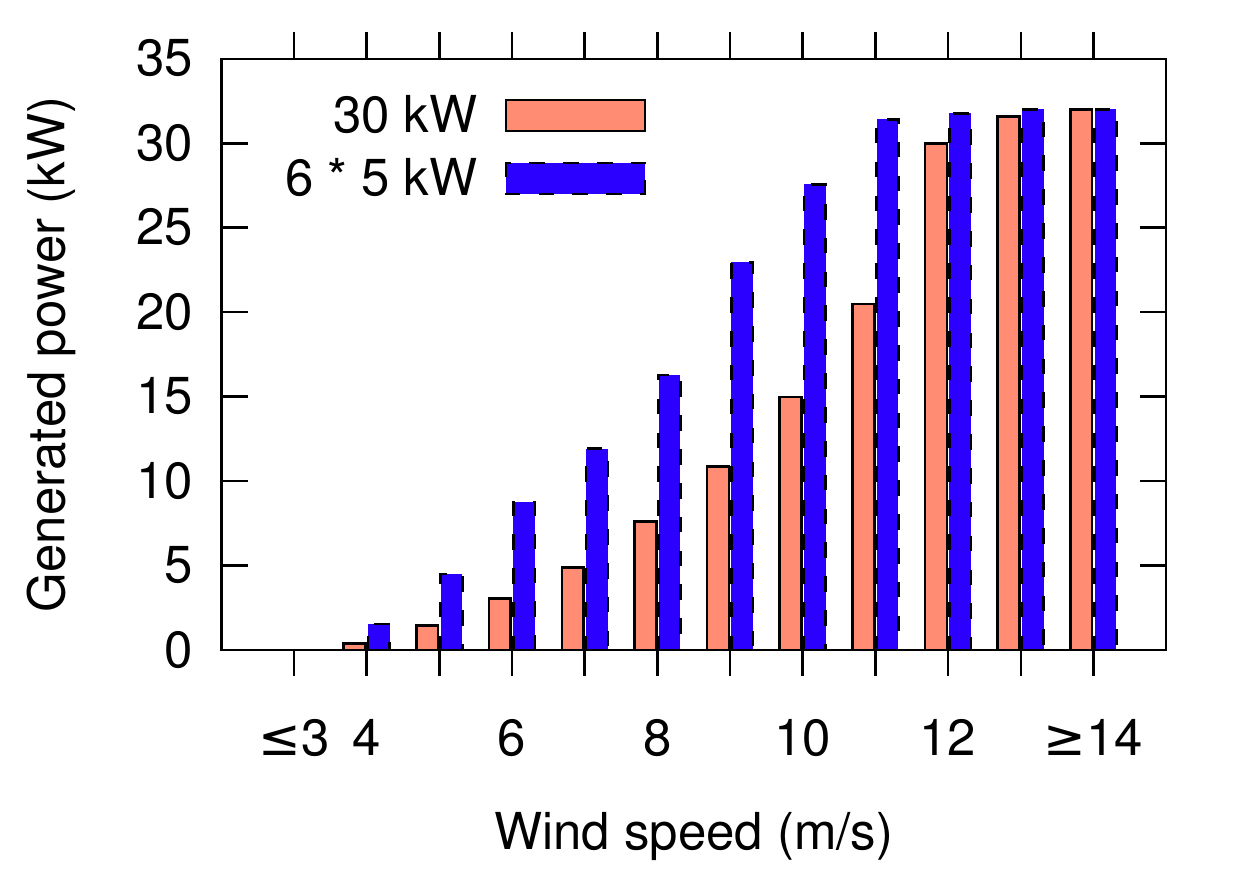}}
  \subfloat[Optimal mix distribution]{\label{fig:diffDistrib}\includegraphics[width=.5\figurewidth]{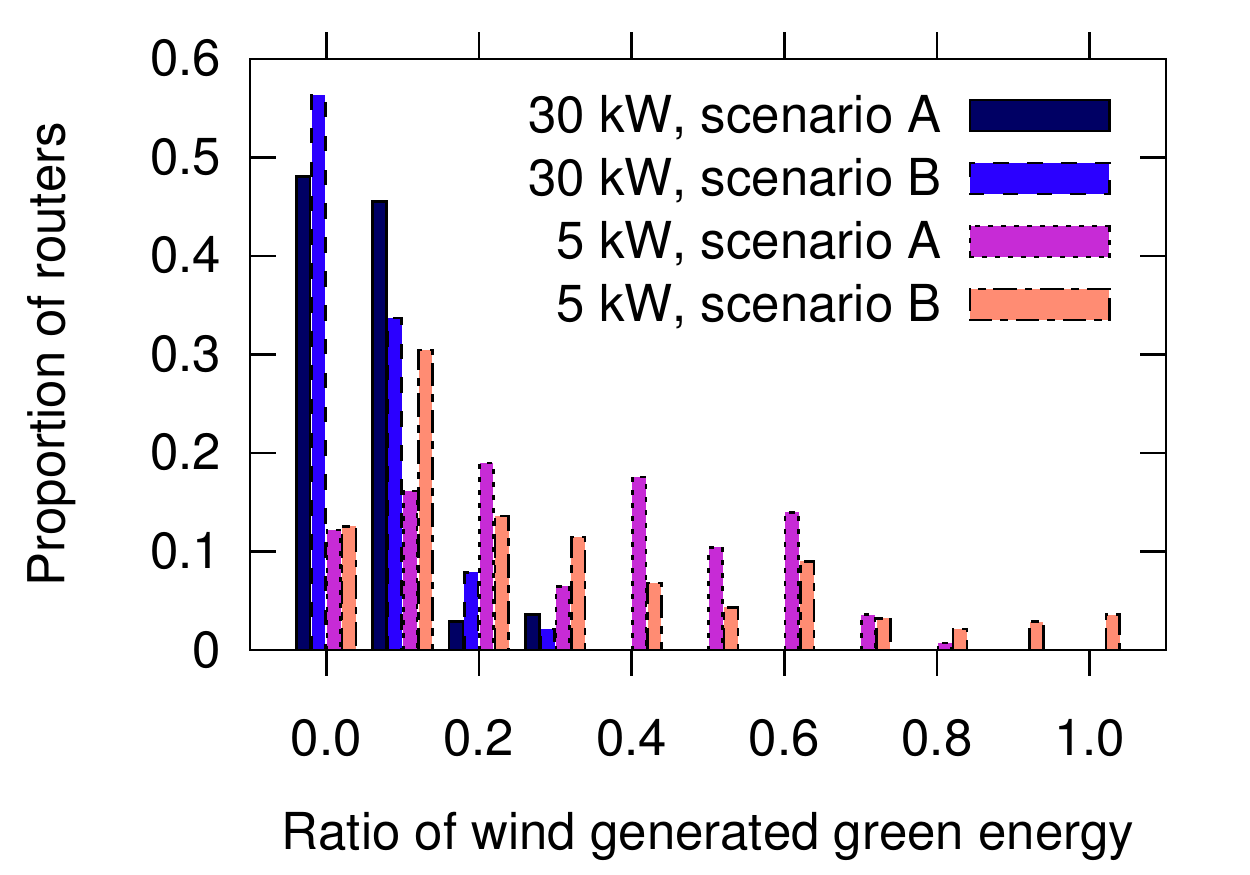}}
 \caption{Wind energy profile for different turbine types (\kW{5} and \kW{30}), as well as distribution of routers with optimal ratio of wind and solar energy sources.}
 \label{fig:impact}
\end{figure}

\figurename{~\ref{fig:windToPower}} presents a plot showing the
relationship of wind speed to generated power that can be achieved by
the two different turbines. As we can see, six \kW{5} wind turbines
produce more power than a single \kW{30} turbine at a low wind speed
(e.g. at $6~m/s$, the ratio is 2.9). However, the type of turbines could be decided on the space restrictions and the cost of the infrastructure.

\subsection*{Solar energy profile}

The Global Horizontal Irradiance (GHI) is used to estimate how much
power could be generated by photovoltaic solar panels. A \kW{4} photovoltaic solar panel that is needed to produce \kW{4} of power,
will require a GHI of $1000~W/m^2$. As shown in
\figurename {\ref{fig:ghi}}, the GHI is usually stronger in the south
west of the U.S. For example, in the brigther zone (very high GHI), the estimated
energy is greater than $6000~kWh$, while the estimated energy is only $3000~kWh$ in the zone with the lowest GHI.

\subsection*{Renewable energy calculation}

\begin{figure*}[!t]
  \centering
  \subfloat[San Jose, CA - GHI]{\label{fig:powerAndData2Days_sanJose_ghi}\includegraphics[width=.25\linewidth]{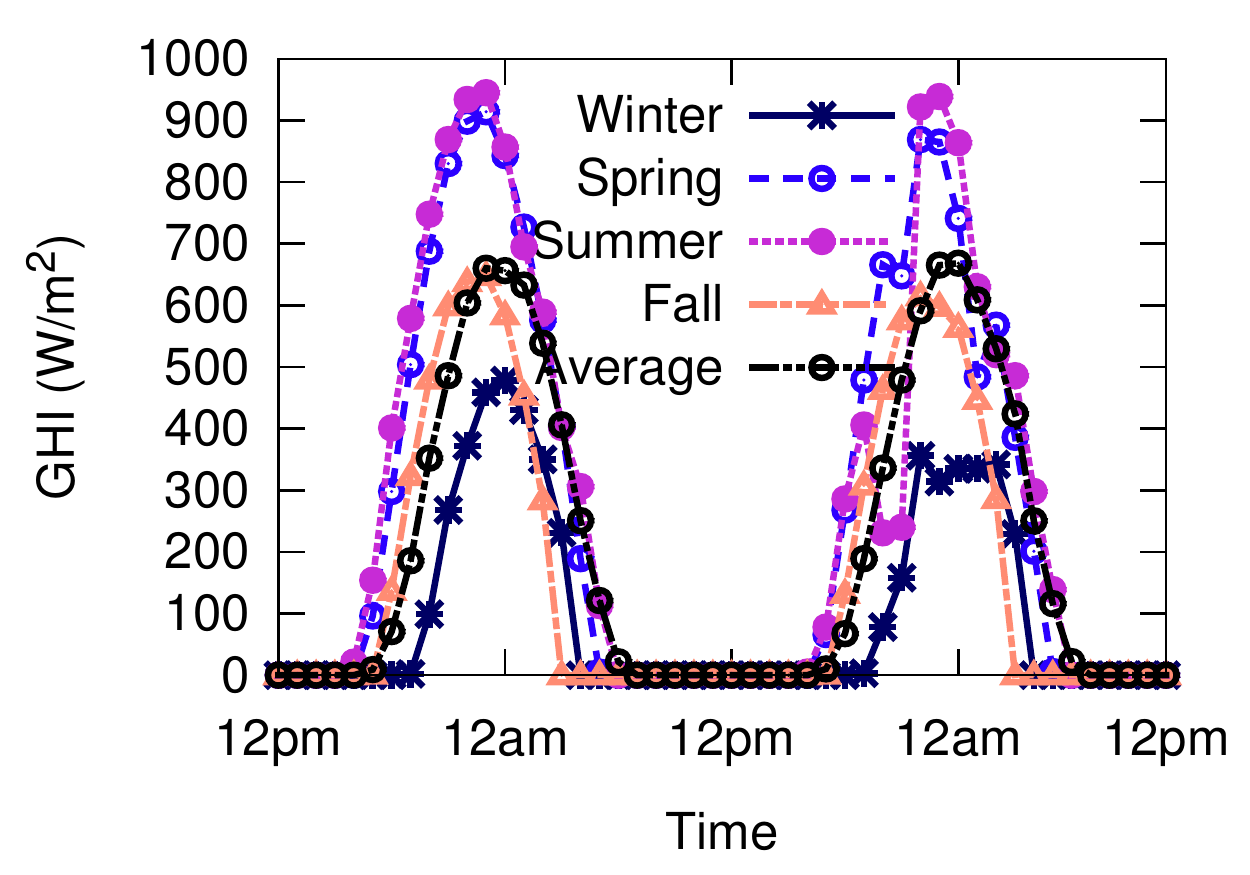}}
  \subfloat[San Jose, CA - Wind speed]{\label{fig:powerAndData2Days_sanJose_windspeed}\includegraphics[width=.25\linewidth]{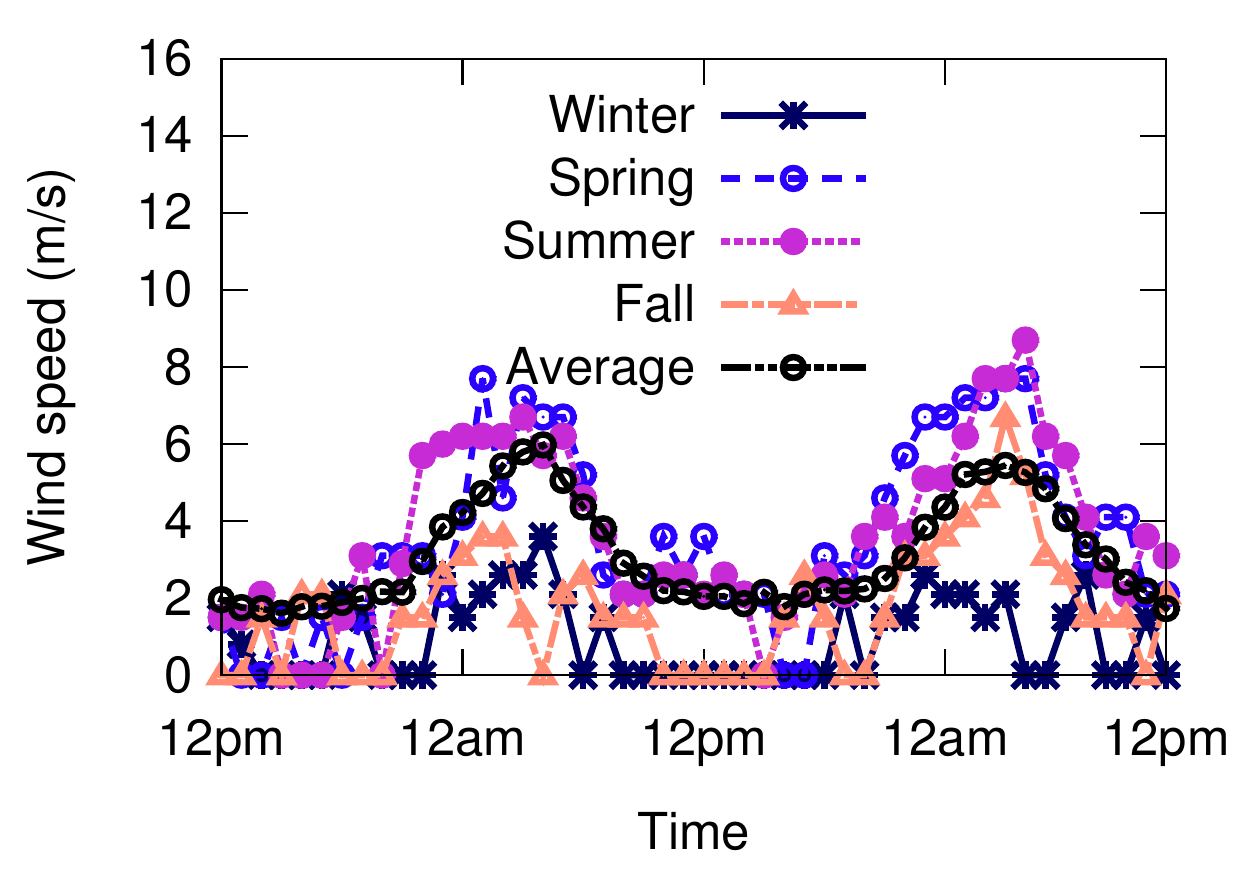}}
  \subfloat[NYC, NY - GHI]{\label{fig:powerAndData2Days_newYork_ghi}\includegraphics[width=.25\linewidth]{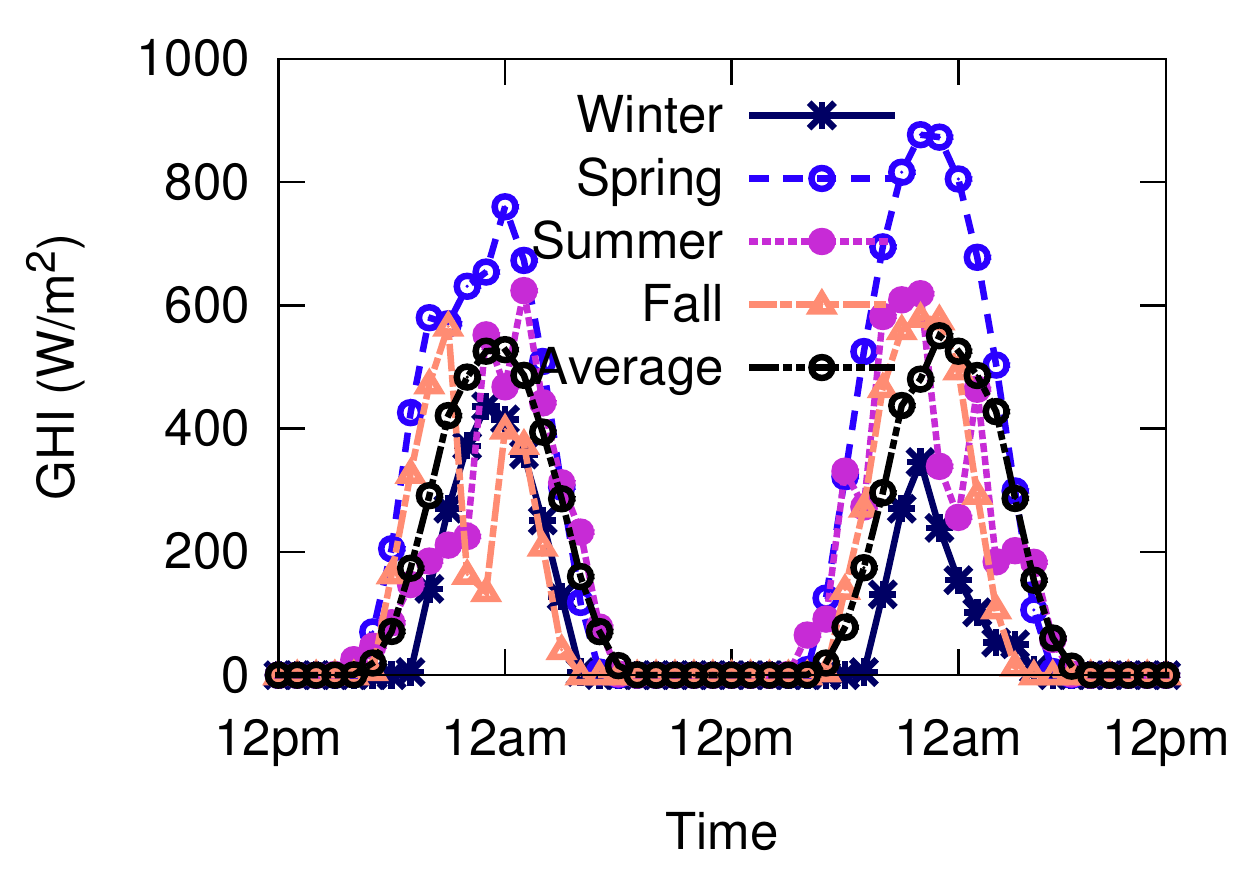}}
  \subfloat[NYC, NY - Wind speed]{\label{fig:powerAndData2Days_newYork_windspeed}\includegraphics[width=.25\linewidth]{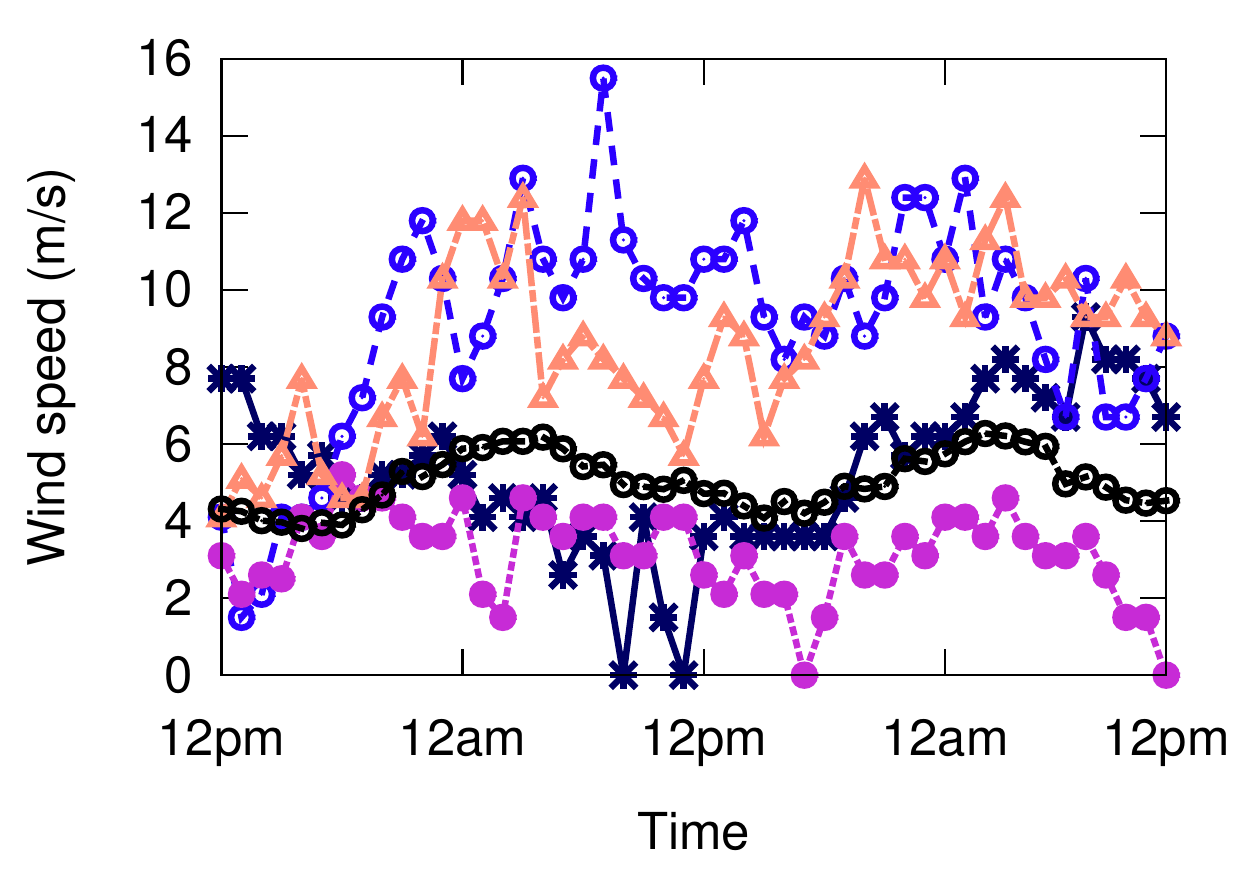}}\\
  \captionsetup[subfloat]{width=.22\linewidth}
  \subfloat[San Jose, CA - Percentage of power generated (with \kW{5} turbine)]{\label{fig:powerAndData2Days_sanJose_5kW_power}\includegraphics[width=.25\linewidth]{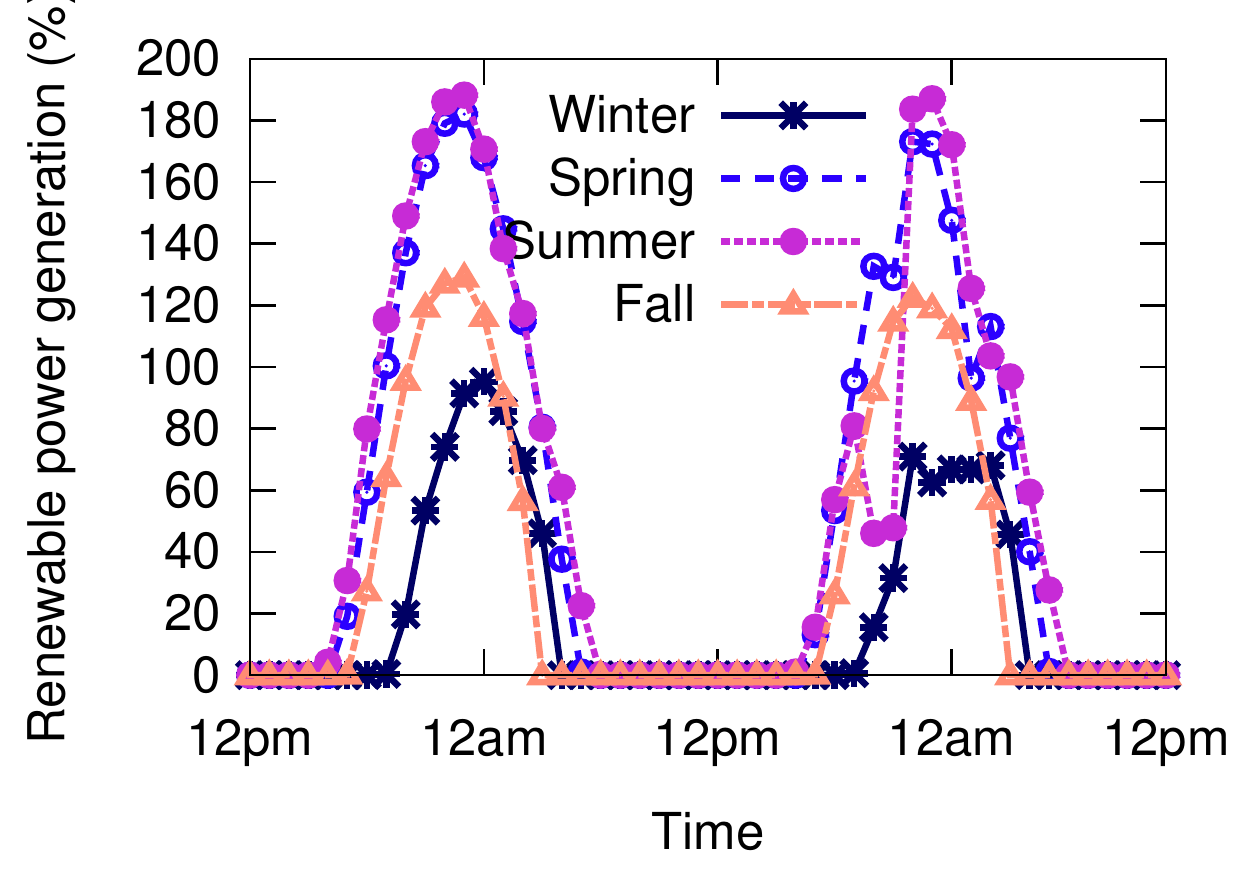}}
  \subfloat[San Jose, CA - Percentage of power generated (with \kW{30} turbine)]{\label{fig:powerAndData2Days_sanJose_30kW_power}\includegraphics[width=.25\linewidth]{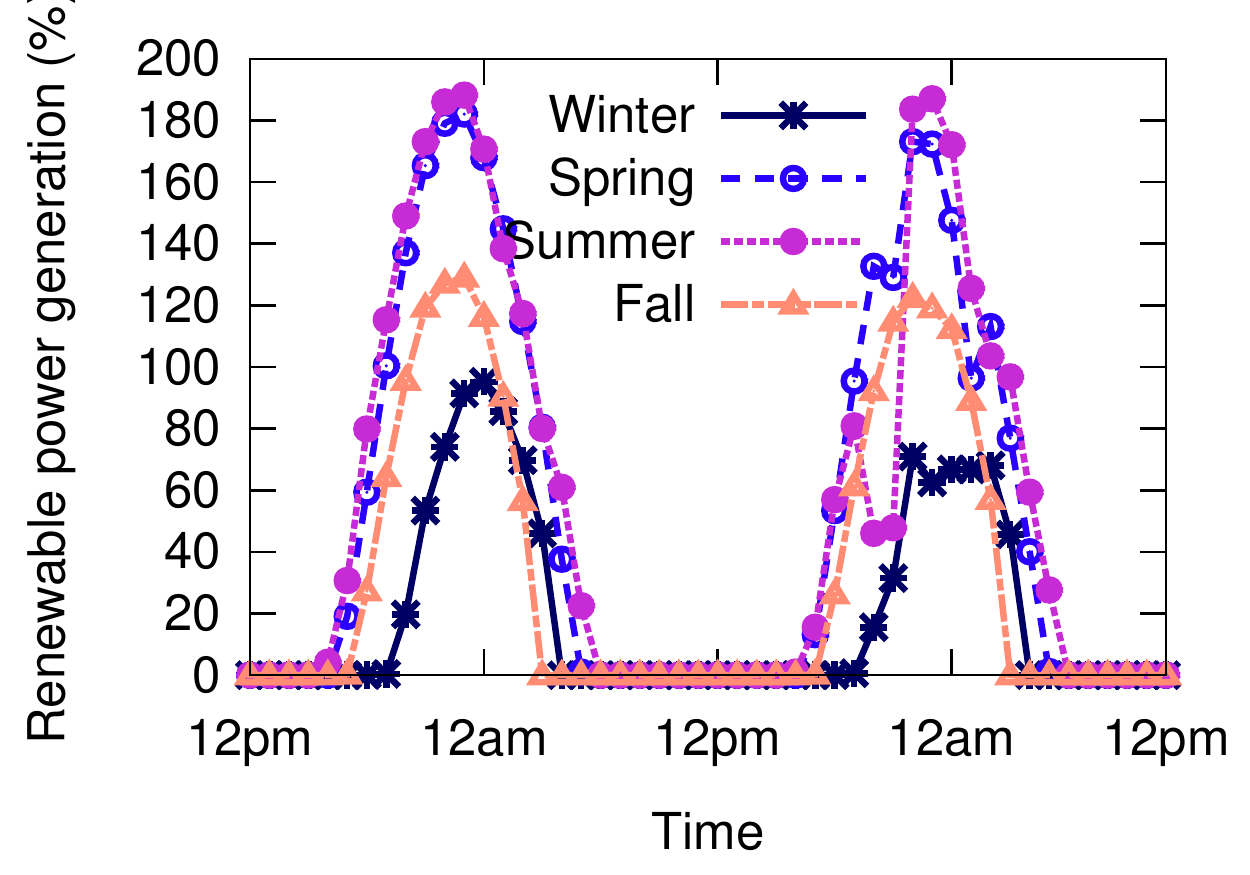}}
  \subfloat[NYC, NY - Percentage of power generated (with \kW{5} turbine)]{\label{fig:powerAndData2Days_newYork_5kW_power}\includegraphics[width=.25\linewidth]{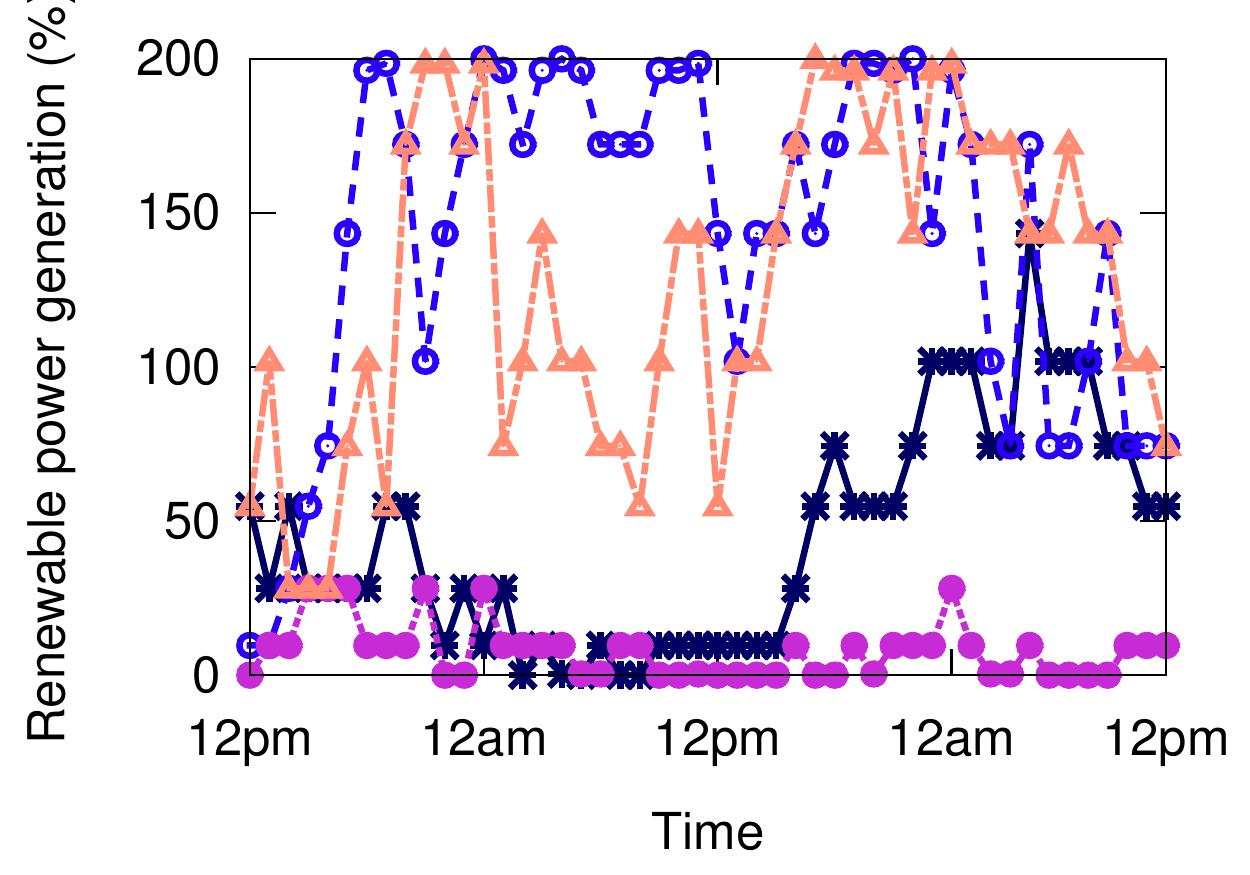}}
  \subfloat[NYC, NY - Percentage of power generated (with \kW{30} turbine)]{\label{fig:powerAndData2Days_newYork_30kW_power}\includegraphics[width=.25\linewidth]{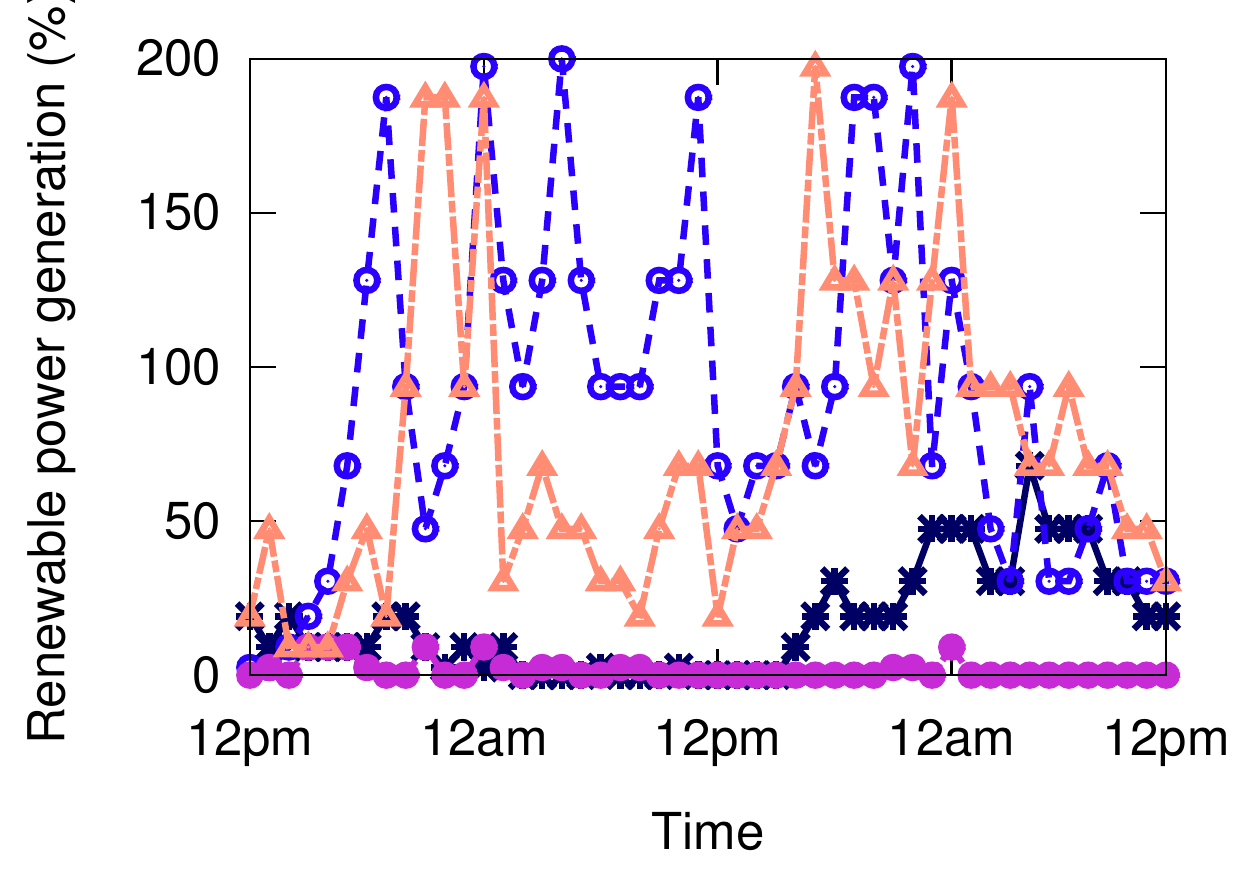}}\\
  \subfloat[San Jose, CA - Green ratio (with \kW{5} turbine)]{\label{fig:powerAndData2Days_sanJose_5kW_greenRatio}\includegraphics[width=.25\linewidth]{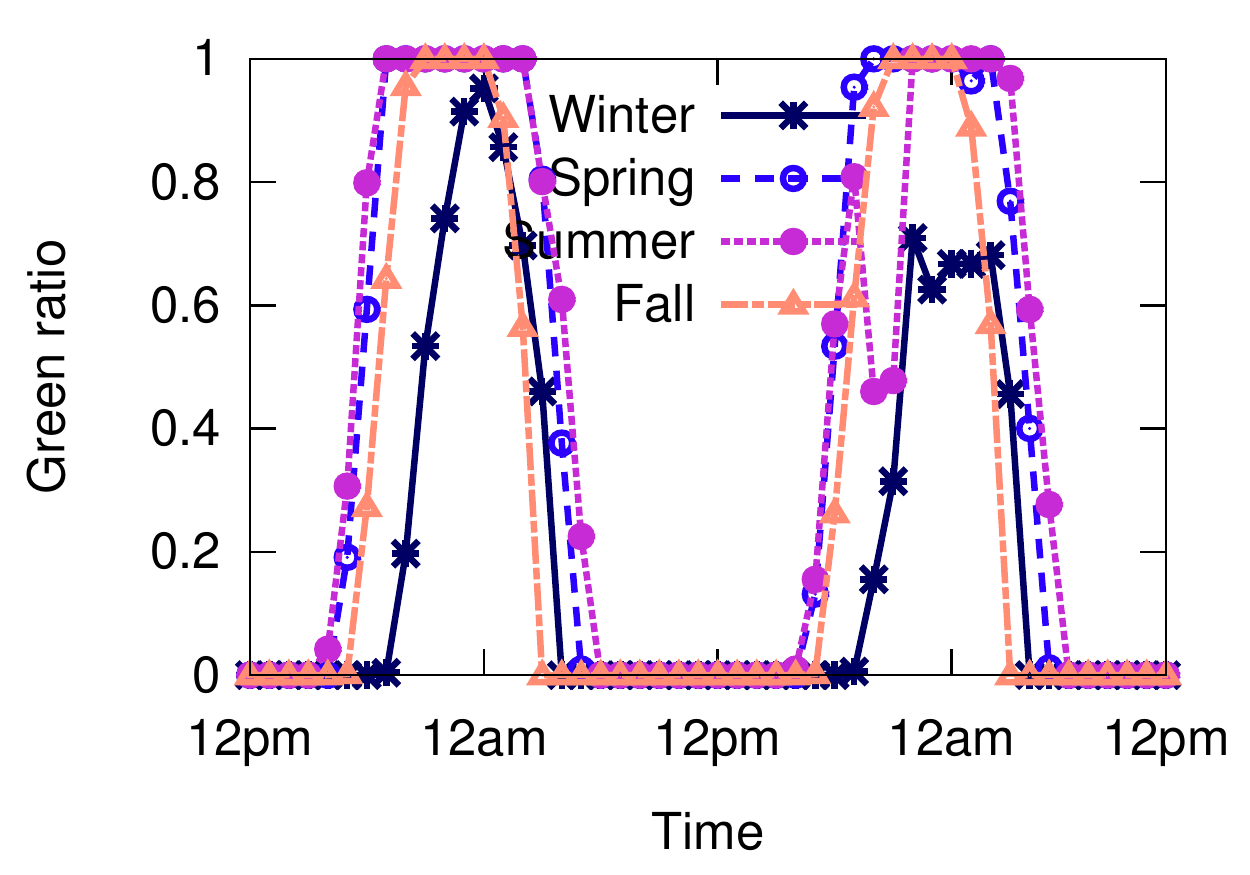}}
  \subfloat[San Jose, CA - Green ratio (with \kW{30} turbine)]{\label{fig:powerAndData2Days_sanJose_30kW_greenRatio}\includegraphics[width=.25\linewidth]{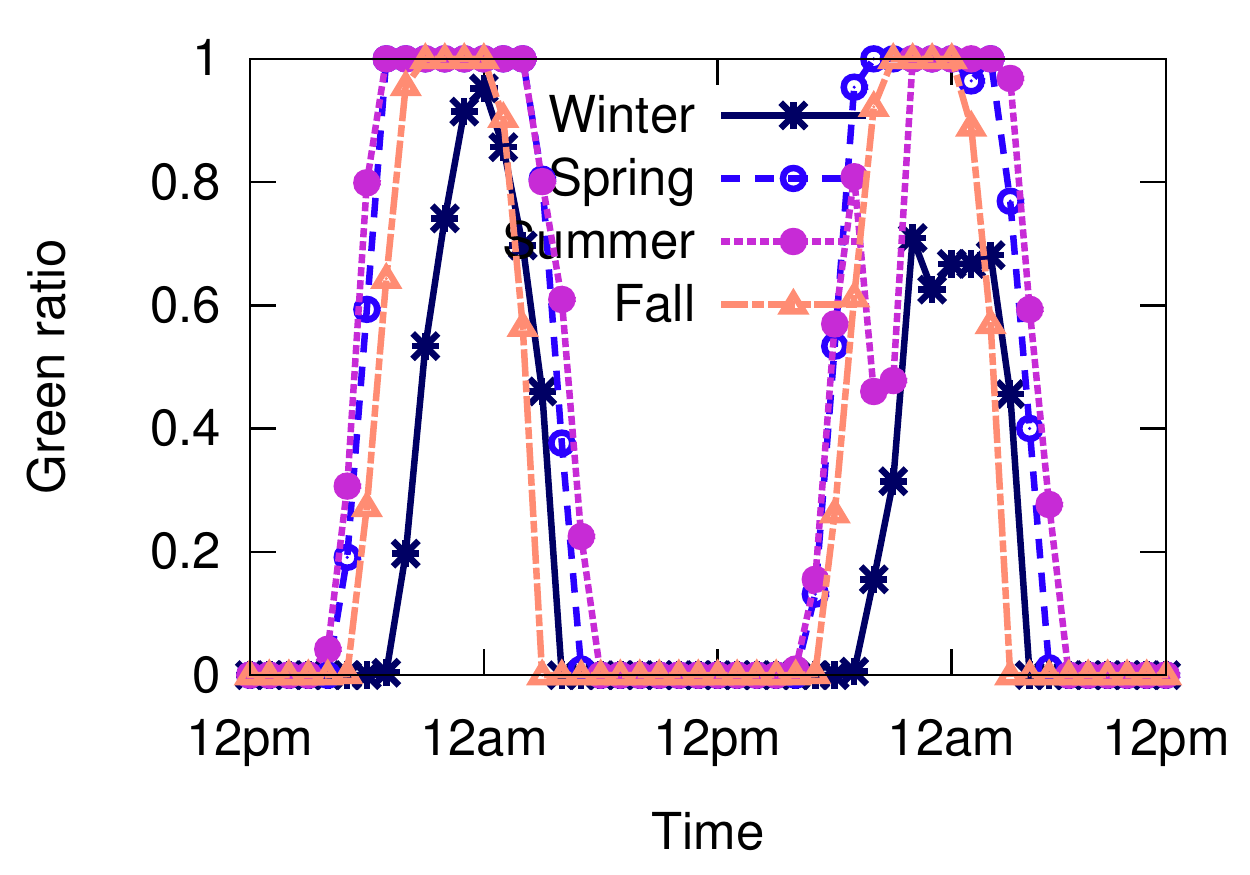}}
  \subfloat[NYC, NY - Green ratio (with \kW{5} turbine)]{\label{fig:powerAndData2Days_newYork_5kW_greenRatio}\includegraphics[width=.25\linewidth]{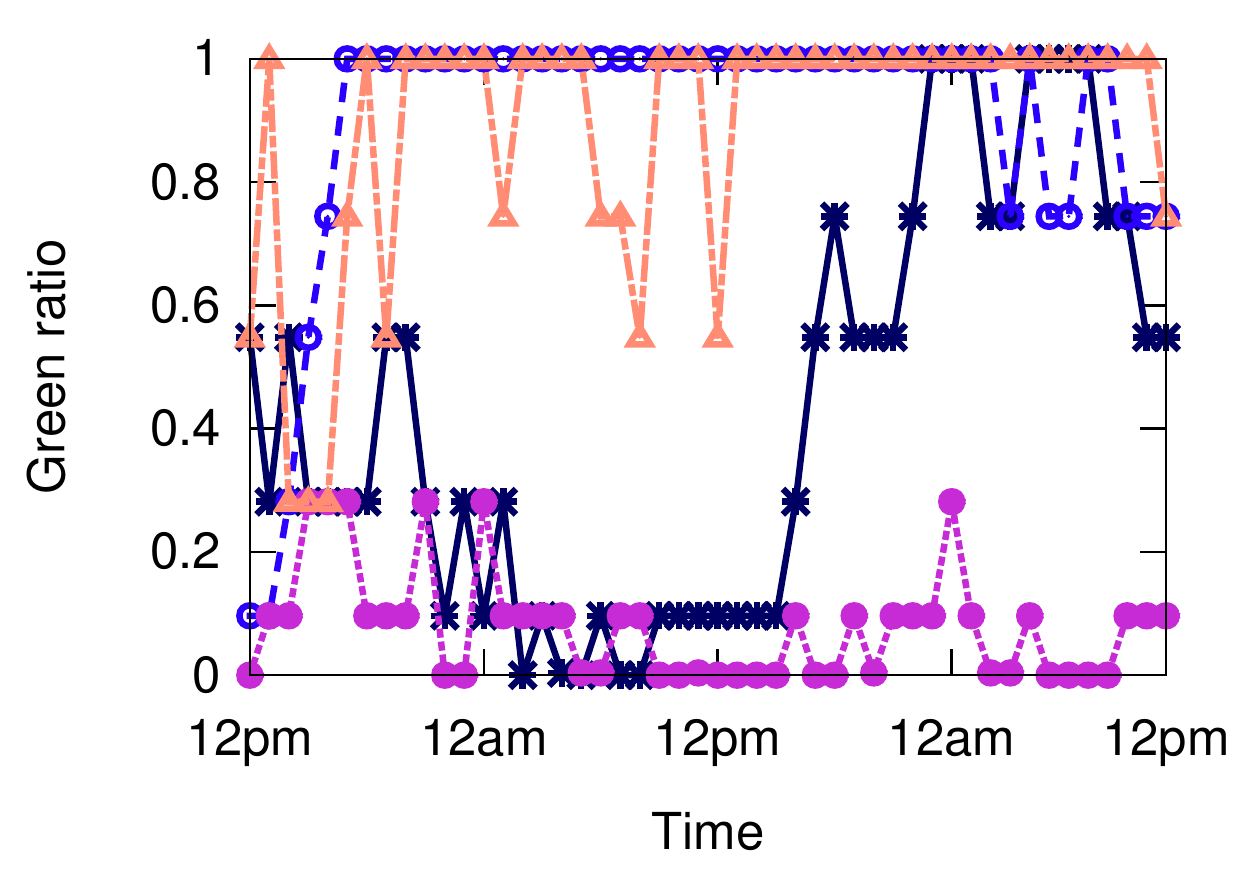}}
  \subfloat[NYC, NY - Green ratio (with \kW{30} turbine)]{\label{fig:powerAndData2Days_newYork_30kW_greenRatio}\includegraphics[width=.25\linewidth]{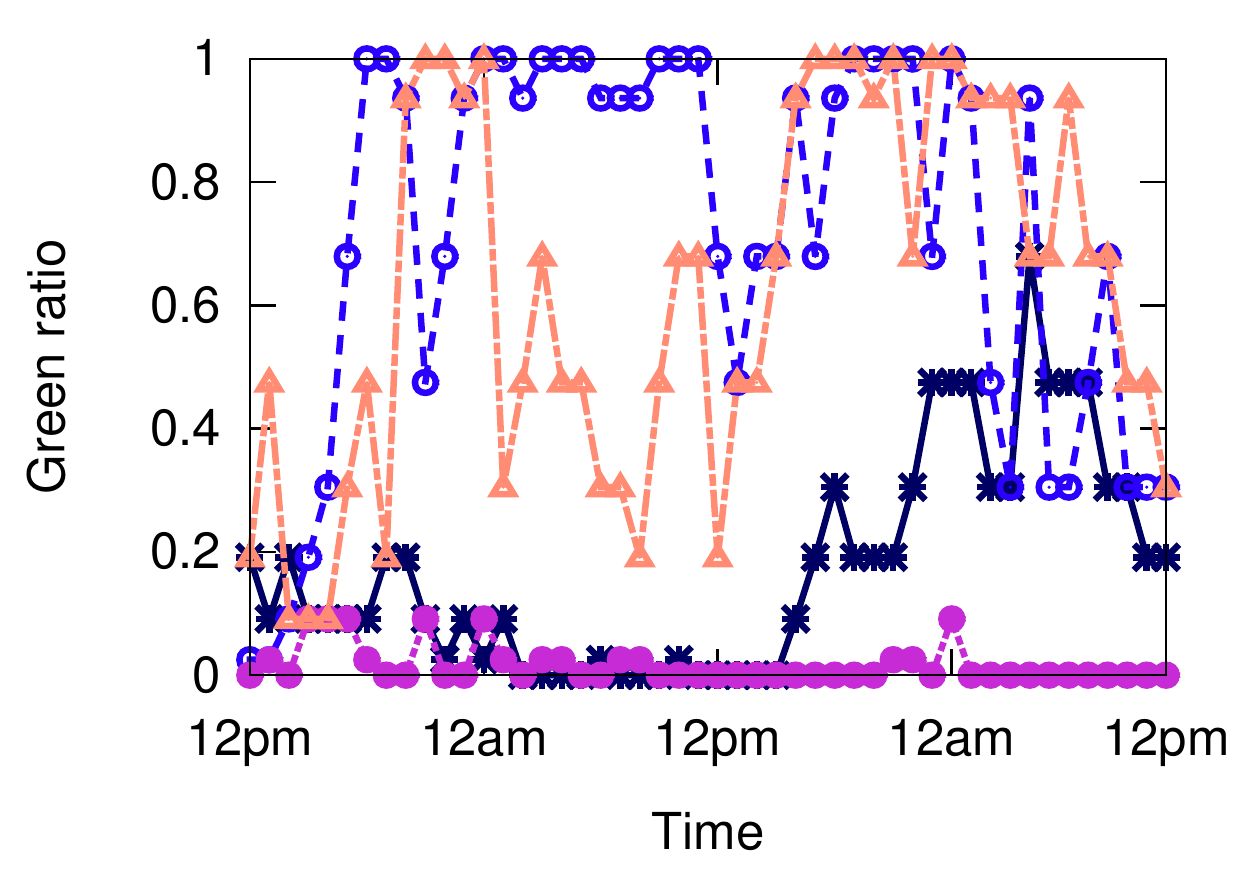}}
  \caption{Weather profiles and power generation performance for San Jose, CA and New York City, NY (The number of turbines has been optimized by calculation \ref{mydef2}).}
  \label{fig:powerAndData}
\end{figure*}

In this section, we describe two calculations that determine the required renewable energy infrastructure to provide adequate renewable energy (the infrastructure refers to the 
number of turbines or size of photovoltaic panels). Calculation 1 will be used when the proportion between wind and solar power is predefined. 
Calculation \ref{mydef2} will determine the optimal proportion of wind and solar power that maximizes the green ratio of the router.
\begin{mydef}
\label{mydef1}
 Given a proportion $\beta$, a capacity $c$ and $PC(X^0,t)$ the energy required by the router $X$, we intend to determine the renewable energy infrastructure, such that: 
\begin{multline}
 rePC(X,t) = P_w(X,t) + P_s(X,t),\\
 \text{ where } P_w(X,t) = \beta rePC(X,t)
\end{multline}

In this case, $P_w(X,t)$ is the energy provided from wind infrastructure, while $P_s(X,t)$ is the energy provided by the solar panels. Therefore, $rePC(X,t)$ 
is the total renewable energy that can be provided to router $X$. 
At the same time we also consider cases when a large quantity of renewable energy is produced. In such situation, the capacity $c$ is set to satisfy $max(rePC(X,t)), \forall t = c PC(X^0,t)$.
\end{mydef}

\begin{mydef}
 \label{mydef2}
 In order to maximize the green ratio of the router, we decided to find the optimal $\beta$
 that would maximize the average $g(X,t)$ ($avg(g(X,t))$), for a typical year. The $avg(g(X,t))$ is calculated as follow:
\begin{equation}
avg(g(X,t)) = \frac{\sum_{t=1}^{24 *365} g(X,t)}{24 *365}
\end{equation} where $g(X,t)$ is the green ratio of router $X$ at time $t$ defined by equation~\ref{eq:green_ratio} 
(time has been shifted to match the PDT time zone of Los Angeles, CA). We will do this calculation for 
all $\beta$ between 0.0 to 1.0. Therefore, the optimal $\beta$ will be the value with the best $avg(g(X,t))$.
\end{mydef}

\figurename{~\ref{fig:diffDistrib}} shows the distribution of routers for the Sprint network topology as a function of the optimal $\beta$. For example, in \figurename{~\ref{fig:diffDistrib}}, when 
no wind energy are considered in the ratio (i.e. $\beta=0.0$), for a maximum $c = 2$ for \kW{30} wind turbine, we can see that only 47\% of the routers will have this configuration. 
\figurename{~\ref{fig:diffDistrib}} shows the optimal mix for both types of turbines for two different types of capacities. The two different capacities, includes a fixed capacity $c = 2$ 
(also referred to as constant capacity for the scenario A) and random capacity that is uniformly selected between 0 and 3 for the scenario B. 
As shown in \figurename{ \ref{fig:diffDistrib}, selecting the \kW{5} turbine will lead to 
a better distribution of routers with mixed sources of wind and solar (where $\beta$ goes up to $1.0$). On the other hand, the \kW{30} turbine will only lead to low ratios of wind energy 
(e.g. $0.0 \leq \beta \leq 0.3$).

We now show an example of the optimization process for two routers,
one in San Jose, CA, and the second in New York City, NY.
\figurename{~\ref{fig:powerAndData2Days_sanJose_ghi} and \ref{fig:powerAndData2Days_newYork_ghi}} show the
GHI and wind speeds for these three locations over a period of two days, for each season.
\figurename{~\ref{fig:powerAndData2Days_sanJose_5kW_power}, \ref{fig:powerAndData2Days_sanJose_30kW_power}, 
\ref{fig:powerAndData2Days_newYork_5kW_power} and \ref{fig:powerAndData2Days_newYork_30kW_power}} show the
power generated at these two locations for wind energy using either \kW{5} or \kW{30}
turbines, after the optimization of ratio between wind and solar energy.
Recall that the optimization is performed over the whole year by picking out the peak periods of high wind and solar performance, and the
figures presented only show performance for two days. Interestingly, the
router in San Jose gets all of its renewables from solar and ignores
the weak wind performance, even though this means that during night time the
router must be powered with brown energy. In contrast, the router in
New York uses only wind energy as its renewable source (although not
clearly visible in \figurename{~\ref{fig:windSpeed}}, New York is in the
medium wind zone).
\figurename{~\ref{fig:powerAndData2Days_newYork_5kW_power} and \ref{fig:powerAndData2Days_newYork_30kW_power}},
shows that using \kW{5} wind turbines produces
better energy performance than \kW{30} wind turbines. As a consequence, only experiments using \kW{5} wind turbines have been presented in this paper.